\documentclass[sigconf]{acmart}

\usepackage{wrapfig}
\usepackage{float}
\usepackage{soul}
\usepackage{hyperref}

\usepackage{graphicx}
\usepackage{tikz}
\usepackage{color}
\usepackage{subcaption}
\usepackage{booktabs}

\definecolor{mypurple}{RGB}{119, 69, 198}

\newcommand{\knm}[1]{\textcolor{black}{ #1}}

\AtBeginDocument{%
  \providecommand\BibTeX{{%
    \normalfont B\kern-0.5em{\scshape i\kern-0.25em b}\kern-0.8em\TeX}}}

\copyrightyear{2023} 
\acmYear{2023} 
\setcopyright{rightsretained} 
\acmConference[CHI '23]{Proceedings of the 2023 CHI Conference on
Human Factors in Computing Systems}{April 23--28, 2023}{Hamburg,
Germany}
\acmBooktitle{Proceedings of the 2023 CHI Conference on Human Factors
in Computing Systems (CHI '23), April 23--28, 2023, Hamburg,
Germany}\acmDOI{10.1145/3544548.3581481}
\acmISBN{978-1-4503-9421-5/23/04}

\begin{document}

\title{Towards Inclusive Avatars: Disability Representation in Avatar Platforms}


\author{Kelly Mack}
\authornote{Author performed this work while at Snap Inc.}
\affiliation{%
  \institution{Snap Inc. and Paul G. Allen School of Computer Science, University of Washington}
  \city{Seattle}
  \state{WA}
  \country{USA}
}

\author{Rai Ching Ling Hsu}
\affiliation{%
  \institution{Snap Inc.}
  \city{Toronto}
  \state{Ontario}
  \country{Canada}
}

\author{Andr\'es Monroy-Hern\'andez}
\authornotemark[1]
\affiliation{%
  \institution{Snap Inc. and Princeton University}
  \city{Princeton}
  \state{NJ}
  \country{USA}
}

\author{Brian A. Smith}
\authornotemark[1]
\affiliation{%
  \institution{Snap Inc. and Columbia University}
  \city{New York}
  \state{NY}
  \country{USA}
}

\author{Fannie Liu}
\authornotemark[1]
\authornote{This paper was prepared while the last author was at Snap Inc. and does not have contributions from the Global Technology Applied Research center of JPMorgan Chase \& Co. This paper is not a product of the Research Department of JPMorgan Chase \& Co. or its affiliates. Neither JPMorgan Chase \& Co. nor any of its affiliates makes any explicit or implied representation or warranty and none of them accept any liability in connection with this paper, including, without limitation, with respect to the completeness, accuracy, or reliability of the information contained herein and the potential legal, compliance, tax, or accounting effects thereof. This document is not intended as investment research or investment advice, or as a recommendation, offer, or solicitation for the purchase or sale of any security, financial instrument, financial product or service, or to be used in any way for evaluating the merits of participating in any transaction.}
\affiliation{%
  \institution{Snap Inc. and JPMorgan Chase}
  \city{New York}
  \state{NY}
  \country{USA}
}

\renewcommand{\shortauthors}{Mack et al.}

\begin{abstract}
    Digital avatars are an important part of identity representation, but there is little work on understanding how to represent disability. We interviewed 18 people with disabilities and related identities about their experiences and preferences in representing their identities with avatars. Participants generally preferred to represent their disability identity if the context felt safe and platforms supported their expression, as it was important for feeling authentically represented. They also utilized avatars in strategic ways: as a means to signal and disclose current abilities, access needs, and to raise awareness. Some participants even found avatars to be a more accessible way to communicate than alternatives. We discuss how avatars can support disability identity representation because of their easily customizable format that is not strictly tied to reality. We conclude with design recommendations for creating platforms that better support people in representing their disability and other minoritized identities.
\end{abstract}




\maketitle

\section{Introduction}

\knm{People use avatars to represent themselves and express their identity in online environments, ranging from messaging apps to virtual reality (VR), i.e., the metaverse. However, options to represent disability in avatars are often limited or fully missing on popular platforms. Disability affects billions of people and can be a core, positive part of identity \cite{shakespeare1996disability}. When excluded from the design process, people with disabilities can be left feeling neglected and invisible as virtual worlds become more commonplace.} While prior work has studied how people portray and describe disability and other identity characteristics in pictures or stock photos \cite{edwards2021s, bennett2021s}, as well as other identities like gender in video game contexts \cite{mcarthur2015avatar, passmore2018about, morgan2020role}, people with disabilities are often left out of avatar research \cite{mott2019accessible}. One recent work focuses on blind or low vision (BLV) and d/Deaf or Hard of Hearing (DHH) people's experiences with avatars in VR~\cite{zhang2022s}, but more research is needed to understand how those findings extend to social applications more broadly and with other disabilities, including invisible disabilities.  
We investigate how people with disabilities express and \textit{want} to express their disabled identities digitally with avatars.

To study how people with disabilities use and wish to use avatars to represent their identities, we interviewed 18 people who identified as disabled, chronically ill, mentally ill, neurodiverse, and/or fat. \knm{While not everyone in these groups may identify as disabled, they share similar experiences of moving through the world with non-normative bodies and minds, which are relevant to making avatars more inclusive.} We recruited people with a diverse range of ability statuses (including people with multiple disabilities), race, gender, and sexuality. More specifically, we investigated the research questions: (1) How do disabled people represent disability in avatars today? (2) How can avatars better represent disabled peoples' experiences and identities? And (3) When and why do people share disability identities or experiences using avatars?

Our findings focus on how context can affect people's desire to represent their disability in their avatar, as well as how avatars can be more accessible than other forms of communication.
Further, we highlight the different customizable components of avatars and how each could be used to represent disabilities and disabled experiences.
Finally, participants discussed how norms and beliefs from both the physical world and avatar platforms shape each other, including the negative impact of biases and stereotypes that persist in avatars. Overall, we find many opportunities for avatars to help disabled people represent and even affirm their identities---provided that avatar platforms understand and support their needs.
 
In summary, this work contributes:
\begin{itemize}
    \item Insights into how avatars function as a form of identity expression, including for people with invisible or fluctuating disabilities
    \item Examples of how avatars can provide accessible ways of communicating, especially about disability
    \item A characterization of how different dimensions of avatars can support disability representations
    \item A list of design considerations for building avatar platforms that support disability representation
\end{itemize}

\section{Background and Related Work}

\knm{We first provide background on two central topics of this paper: disability and digital avatars. We then discuss identity representation in avatars and its fluctuation, specifically focusing on the affordances to people with minoritized identities like disability.}

\subsection{The Social Model of Disability}

\knm{While there are several models of disability \cite{cobley2018understanding}, we adopt the social model of disability that views people as disabled not because of their physical impairments, but because of social barriers that were created by a world that did not consider all bodies and minds and variations in ability \cite{shakespeare2006social, goering2015rethinking, barnes2019understanding}. Oftentimes HCI accessibility research focuses primarily on people with sensory-related disabilities (e.g., blind, mobility impairment, deaf) \cite{mack2021we}. However, in this work, we seek to include other, less considered groups, such as people who are neurodiverse or chronically ill, as they all often share experiences of facing social barriers because of these identities. Adopting a social model view of disability allows for a positive view of disability identity based on both personal experience and the disability community's shared experiences, culture, and political agendas \cite{shakespeare1996disability}. Therefore, regarding avatars, taking the lens of the social model of disability leaves space for identity pride and community to be at the forefront of disability representation in avatars.}

\subsection{Introduction to Digital Avatars}

\begin{figure}
\captionsetup{justification=centering}
\begin{subfigure}{.45\textwidth}
  \centering
  \includegraphics[height=1.4in]{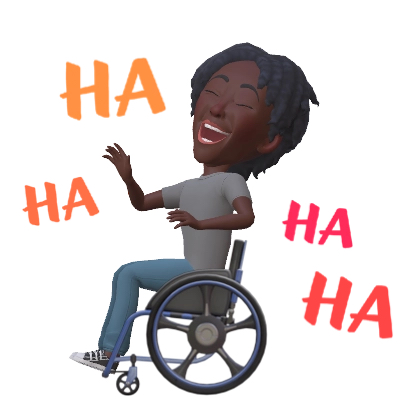}
  \caption{Avatars can incorporate assistive technologies or other objects to express their current mood or situation.}
\end{subfigure}\hfill%
\begin{subfigure}{.45\textwidth}
  \centering
  \includegraphics[height=1.4in]{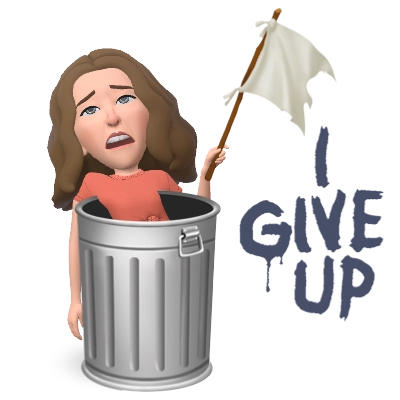}
  \caption{Avatars can incorporate words and backgrounds, often in a cartoon style.}
\end{subfigure}\hfill%
\caption{These Meta avatars demonstrate how one can add cartoon objects and annotations for a message.}
    \label{fig:annotations}
    \Description{Two avatars. The first is a black person with braided black hair in a wheelchair laughing with little bubbles letter "ha's" everywhere. The second is a white person with shoulder-length brown hair slumped over in a trash can waving a white flag with the words "i give up."}
\end{figure}

Prior research defines digital avatars (which we refer to as ``avatars'') in a variety of ways, but most definitions agree that an avatar ``represents the user in a digital environment'' and that the avatar ``enable[s] the user to experience and interact within the spaces of digitally mediated worlds'' \cite{nowak2018avatars}. In this work, we focus specifically on humanoid/human-like avatars. Humanoid avatars can range from static pictures of avatars (e.g., profile pictures on social media) to 3D renderings of humanoids that can interact with their environments (e.g., video game avatars). The customizable features of an avatar vary from platform to platform, but many enable the creator to alter a host of physical characteristics such as hairstyles, facial features, skin tone, height, weight/build, clothing, and poses. 
Other platforms allow customization around the actions the avatar can perform. For example, in \textit{The Sims 4}~\cite{sims4}, and \textit{Grand Theft Auto V}~\cite{gtav}, players can customize how the avatar walks. 
Additionally, some platforms allow users to add backgrounds, words, and objects or props for additional customization (see \autoref{fig:annotations}). We further scope our focus on avatars that can be customized in physical appearance and/or clothing.

People can use avatars in various digital spaces and interactions, from video games to social apps. Several social app avatars are available on their native platforms for messaging purposes. For example, on iMessage, users can record their Memojis making different faces, and Meta, Memoji, and Bitmoji avatars enable users to share stickers (preset scenes or phrases) with their avatars in text messages, social media posts, and email.

\subsection{Avatars and Identity}
In interactions with others, Goffman posits that people manage how they present themselves, trying to balance how they want to represent themselves and how they want others to perceive and interact with them \cite{goffman2021presentation}. More recent work found that even in virtual interactions with avatars, self-presentation is important. When using a customized avatar, users can feel it reflects their identities in the physical world and feel more physiologically aroused  \cite{morgan2020role, bailey2009avatar}. How strongly a person feels connected to their avatar depends on several factors, including social norms and how well the avatar reflects their actual versus ideal self~\cite{nowak2018avatars, jin2009parasocial, jin2009avatars}. \knm{Indeed, avatars present an opportunity to present parts of an ideal or true self that are not visible in the physical world \cite{kafai2010your}.}


\subsubsection{Context Affects Identity Presentation}
Prior work categorized people's goals when making avatars into realistic, idealized, or fantastical/unrealistic versions of oneself~\cite{turkle1994constructions, neustaedter2009presenting}. Choosing from these categories and choosing which parts of identity to highlight or hide often depends on context
\cite{turkle1999cyberspace, bruckman1996gender, haimson2016constructing}. For example, in video games, people are encouraged to express themselves in any way, even if it doesn't align with their physical-world appearance; thus, visibly disabled people often choose to present as nondisabled~\cite{davis2021machine, gerling2016designing, zhang2022s}. On the other hand, in text messages or social media posts, people prefer to closely couple their real life and avatar appearances \cite{zhang2022s, freeman2021body}. \knm{The intended audience of the online persona can further affect choices around presenting oneself (and one's identities), such as work colleagues versus friends \cite{dimicco2007identity}.} Further, when making idealized or unrealistic avatars, users have the opportunity to try on new identities altogether \cite{haimson2016constructing, bruckman1996gender, morgan2020role}, thus providing a venue for experimentation at a low risk compared to the physical world \cite{turkle1999cyberspace}. 

\subsubsection{Minoritized Identities in Avatars}
Prior work in HCI investigating identity and avatars largely found that biased assumptions about users were built into customization platforms \cite{passmore2018about, higgin2009blackless, leonard2006not, pace2009socially, pace2008can, consalvo2003s, mcarthur2014everyone, mcarthur2015avatar,kafai2010blacks}. 
For example, studies found that certain games chose defaults that perpetuated racial stereotypes, such as white skin tones for protagonists and darker skin tones for antagonists \cite{mcarthur2015avatar, leonard2006not}. Moreover, changing skin color is not enough to represent people of color more broadly \cite{mcarthur2015avatar}; shapes of eyes, mouths, hairstyles, and hair textures are also key physical characteristics \cite{passmore2018about}.
Minoritized genders and sexualities have similarly been underrepresented or represented in biased ways (e.g., defaulting to male avatars, hyper-sexualizing female avatars) \cite{mcarthur2014everyone, consalvo2003s, mcarthur2015avatar}. Researchers have suggested different strategies to mitigate bias, including allowing for diverse pronouns \cite{morgan2020role}, allowing for any combination of physical characteristics, avoiding male-female dichotomies \cite{morgan2020role, mcarthur2015avatar}, and avoiding non-randomized defaults \cite{mcarthur2015avatar}. In this work, we explore the biased representations that people with disabilities may experience on avatar platforms.

Finally, Crenshaw's concept of intersectionality theorizes that people's different identities intersect and result in unique consequences \cite{crenshaw1990mapping}. This concept might lead to unique ways of presenting identities—especially disability, ethnicity, race, and gender—in an avatar \cite{edwards2020three}. In this work, we specifically recruited people with diverse identities to understand the unique needs of disabled people of color and queer disabled people.

\subsubsection{Disability Identity Representation in Avatars}

There are a few avatar platforms that provide positive examples of identity representation in avatars. For example, concerning disability, \textit{Overwatch} has several characters that use prosthetics, as well as an older adult woman of color with only one eye and another who is neurodiverse. The community response to these characters was largely positive \cite{cullen2018better}, highlighting the meaningfulness of representation in video games. A few social apps also provide assistive technology representation in their avatars. For example, Meta avatars, Memojis, and Bitmojis each allow users to show a hearing aid or cochlear implant \cite{zhang2022s} on the avatar, while Meta and Bitmoji also support wheelchairs in their avatar stickers. 

Disability representation, especially outside of video games, is conspicuously absent from existing literature \cite{mott2019accessible}. One exception is Zhang et al.'s work, which focuses on how and why BLV and DHH people represent their disabilities in VR spaces \cite{zhang2022s}. They found that many people use assistive technologies to visually signify their disabilities in avatars. While their participants often aimed to represent their physical selves, sometimes they chose to omit their disability identity in certain contexts (e.g., interacting with strangers) or because they wanted to show other parts of their identity.
We expand on these findings to consider other groups commonly left out of accessibility work (e.g., people with neurodivergence, mental illness, chronic illness) \cite{mack2021we}. Several of these identities are considered ``invisible disabilities,'' which may have different considerations for disability representation than prior work. We also discuss both the accessibility barriers and affordances that avatars provide disabled users.

\section{Methods}
We conducted interviews with 18 people across 2 months. We recruited participants on social media and via disability-focused email lists, using a screening survey that included questions about their demographics, disability, and avatar use.
\subsection{Participants}
\begin{figure}
    \centering
    \includegraphics[width=.9\linewidth]{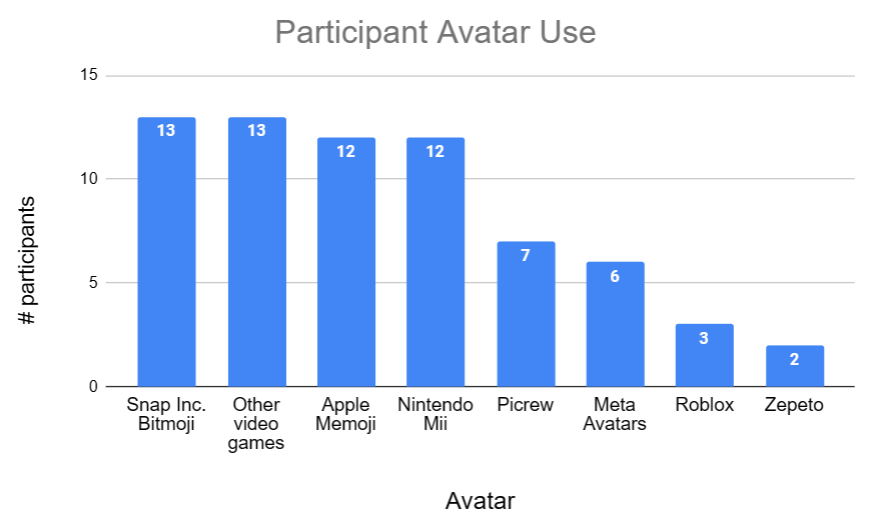}
    \caption{The most used avatars were Bitmojis, video games generally, Memojis, and Miis. Generally, people made avatars that represented themselves more in video game than non-video game contexts.}
    \label{fig:avatar_use}
    \Description{A chart with the number of participants who used each avatar. 13 used Bitmoji, 13 used other video games, 12 used Memoji, 12 used Miis, 7 used Picrew, 6 used Meta, 3 used Roblox, and 2 used Zepeto.}
\end{figure}

\knm{Participants expressed interest in participating in the study by filling out a screening form.} In selecting participants, we first prioritized diversity in disability, then in other demographics. Participants' average age was 29.1 (\textit{SD}=10.5), and 7 as men or trans masc, 6 participants identified as women, and 5 as nonbinary or agender. Concerning race, 9 participants identified as White, 5 as Asian or South Asian, 4 as Black or African American, 2 as Latinx or Hispanic, 1 as other, and 3 as multiple/mixed race. Our participants' avatar platforms are shown in \autoref{fig:avatar_use}.

We recruited broadly for people who identified as disabled, including disabilities like blindness and motor impairments, while also intentionally recruiting for other, often invisible or stigmatized conditions/identities, such as people who identified as chronically ill, mentally ill/having mental health conditions, and being neurodiverse.
\knm{People from these groups do sometimes identify as disabled and often share similar experiences to people who identify as disabled (e.g., neurodiversity \cite{chapman2020neurodiversity}, chronic illness \cite{wendell2001unhealthy}). Additionally, two participants identified as being fat, a dimension of identity that can also lead to similar experiences as disabled people \cite{herndon2002disparate, kai2009fatness}. Though not all people who identify as fat identify as disabled \cite{kai2009fatness}, both of our fat participants identified as disabled, and one of our fat participants explicitly commented that she wanted her fatness to be considered, since it can be viewed as disabling under a social model of disability.} Given our focus on avatar representation for people minoritized because of their bodies or minds, we decided all of these voices were in scope for our work.
Our participant sample included 6 people who are neurodiverse, 6 people with chronic illnesses, 4 DHH people, 4 people with mental illnesses, 3 people with mobility disabilities, 3 BLV people, 2 people who identify as being fat, and 2 people with other disabilities; 10 out of 18 participants had multiple disabilities/conditions.

\subsection{Procedure}
Participants completed a 60-minute semi-structured video call interview and were compensated with a \$50 gift card or donation to charity. \knm{Before the interview, participants filled out a digital consent form to participate in the study and have their interview recorded. At the start of the interview, we reconfirmed consent to record and explained that participants could stop the interview at any time and choose not to answer any questions.} Following the guidance of Mack et al. \cite{mack2022anticipate}, we worked with each participant to ensure the interview was accessible to them, for instance, by hiring captioners, ensuring the interviewer's face was clear for lip reading, and sending questions ahead of time. We started the interview by defining identity: ``a person's sense of self, established by their unique characteristics, affiliations, and social roles.'' Participants shared their experience with disability and other relevant identities that might affect how they represent themselves in avatars. We also showed screenshots of participants' avatars, and asked about their current avatar use, including how and where they use the avatar, why they made the personalization choices they did, and how the avatar does (or does not) reflect their identities. We then asked questions about participants' ideal avatar representation. Participants indicated which avatar they felt represented them the best, explained why, discussed how they would further change or improve the avatar, and how contextual factors might influence their use of these avatars.

\subsection{Analysis}
\knm{We performed reflexive thematic analysis on the interview transcripts with our research questions as guiding inquiries \cite{braun2006using, braun2019reflecting}. Our process was inspired by that described by \cite{braun2006using}, where the first author reviewed all of the interviews while making an initial list of codes, repeatedly referencing prior interviews as they developed and refined new codes. Along the way, they created a codebook, which included codes for interesting patterns/facets of the data. They shared the codebook with the full research team, who provided comments based on their expertise in the field and the subset of interview transcripts they had read, and the first author refined the codebook accordingly. Some example codes include safety concerns, showing the current level of ability with avatars, and the inability to represent two identities at once. Once the codebook was finalized, we verified its completeness by having another author apply the codes to a transcript. The first author and this author discussed any differences in the interpretation and application of codes. After iterating on the codebook based on this discussion, the lead author applied the final set of codes to all transcripts. The authors then summarized the codes into higher-level themes, which are presented below. Example themes included contextual factors on representation, methods of representing disability, and representing multiple, fluctuating identities.}

\section{Results}

Participants used avatars from social platforms,  video games, and avatar websites (\autoref{fig:avatar_use}). Most participants shared these avatars primarily with family and friends, though a few people used them at work. Participants described sharing their avatar with others, frequently as posed stickers with text or props (see \autoref{fig:annotations}), as well as for more persistent situations such as one's Zoom picture (P5). 
Across all contexts, participants enjoyed the creativity, fun, and personalization of using avatars to express themselves.

In the following, we detail the key themes elicited by our analysis. First, we describe when and why participants used avatars both as an accessible communication method and to convey their disability identities. Then, we explain how people portray their disabilities in avatars and their desires to go beyond that representation. Finally, we illustrate how the norms of the physical world and virtual spaces shape each other and affect avatar user experiences. \knm{To protect the identities of our participants, we created similar avatars to the ones participants shared in screenshots to use in this paper, changing non-critical elements.}

 \subsection{How, When, and Why Disabled People Use Avatars to Signal Disability}

\knm{Participants made and used avatars for three main reasons: 1) to represent their identities, where participants had two, sometimes conflicting, goals of accurately representing their identities and controlling how others see them \cite{goffman2021presentation}; 2) to convey their fluctuating needs and identities, as avatars have the unique ability to easily show changes to identities and abilities; and 3) because avatars were a more accessible method of communication as a fast, visual way to communicate without using many words.}


\subsubsection{People Incorporated Disability into Their Avatars to Show Representation and Pride}

Participants showed their minoritized identities to reinforce their pride and embrace their authentic selves. P9 explained: 
\begin{quote}
    \textit{``I spent so much time getting comfortable with who I am ... I'm not interested in not being that in every space that I can as soon as possible.''} ---P9.
\end{quote}
After dealing with internalized and externalized biases and stigma, P9 wanted to demonstrate their hard-earned pride in all virtual spaces. Other times, participants wanted to show their identity as a way to signal to other disabled people that they can exist in certain spaces, \knm{and show nondisabled people that they have pride in their identities}:
\begin{quote}
    \textit{``Yes. I'm proud to be a little person too, and I want other little people to know I'm a little person, so they know that they're identified, that they're represented, especially, since I have a more visible role ... [it might] make it easier for people to imagine themselves accomplishing what they want to accomplish in their career,''} ---P5.
\end{quote}
P5, and other participants like P3 and P6, expressed a desire to make others feel less alone and more capable through representation. \knm{Another way people used avatars to change others' perceptions of their identity was through increased awareness and education:}
\begin{quote}
    \textit{``I didn't figure out my neurodivergencies [until I was] 40. Now I see it everywhere in so many people, but I spent so much of my life not understanding why I was so different from everybody ... If there were a way to visually represent it ... to help people understand that, that to me is important, representation is important,''} ---P10.
\end{quote}
In a world that is starting to recognize more differences in human functioning, people sharing their experience with neurodivergence and disability helps others learn about topics that might apply to their own lives. 



\subsubsection{People Shared Disability Identities in Avatars when They Felt Safe}
Most participants aimed to create realistic representations of themselves and their disability identity. However, there were still times when people were less likely to try to build disability into their avatar. First, safety was paramount. Even participants with strong disability pride identified situations where one cannot share an identity without fear of unfair treatment or harm. P6 chose not to show disability when he didn't \textit{``have the bandwidth to deal with the pushback and toxicity that comes from online.''}. Thus, people with disabilities have to weigh the pros of authentic representation with the cost of the hate they might receive. \knm{For participants, these costs only compounded when they had intersectional, minoritized identities (e.g., being black or queer).} Safety is always paramount in deciding how (or if) to reflect a minoritized identity in virtual spaces.


\subsubsection{People Made Trade-Offs around which Identities to Represent in Avatars}

\knm{Participants, who mostly had multiple, intersecting minoritized identities, often described needing to make trade-offs between which identities they represented, mainly due to the platforms they used.} Each avatar platform has different customization options which support different kinds of identity expression, which meant participants could not always express their full identities. P4 explained:

\begin{quote}

  \textit{``Because of some [platform] limitations I have to prioritize identities, because there aren't as many options in a lot of cases, so I have to shuffle some things around. Like I have A and D but not this,''} ---P4

\end{quote}

\begin{figure*}
\captionsetup{justification=centering}
\begin{subfigure}{.45\textwidth}
  \centering
  \includegraphics[height=1.8in]{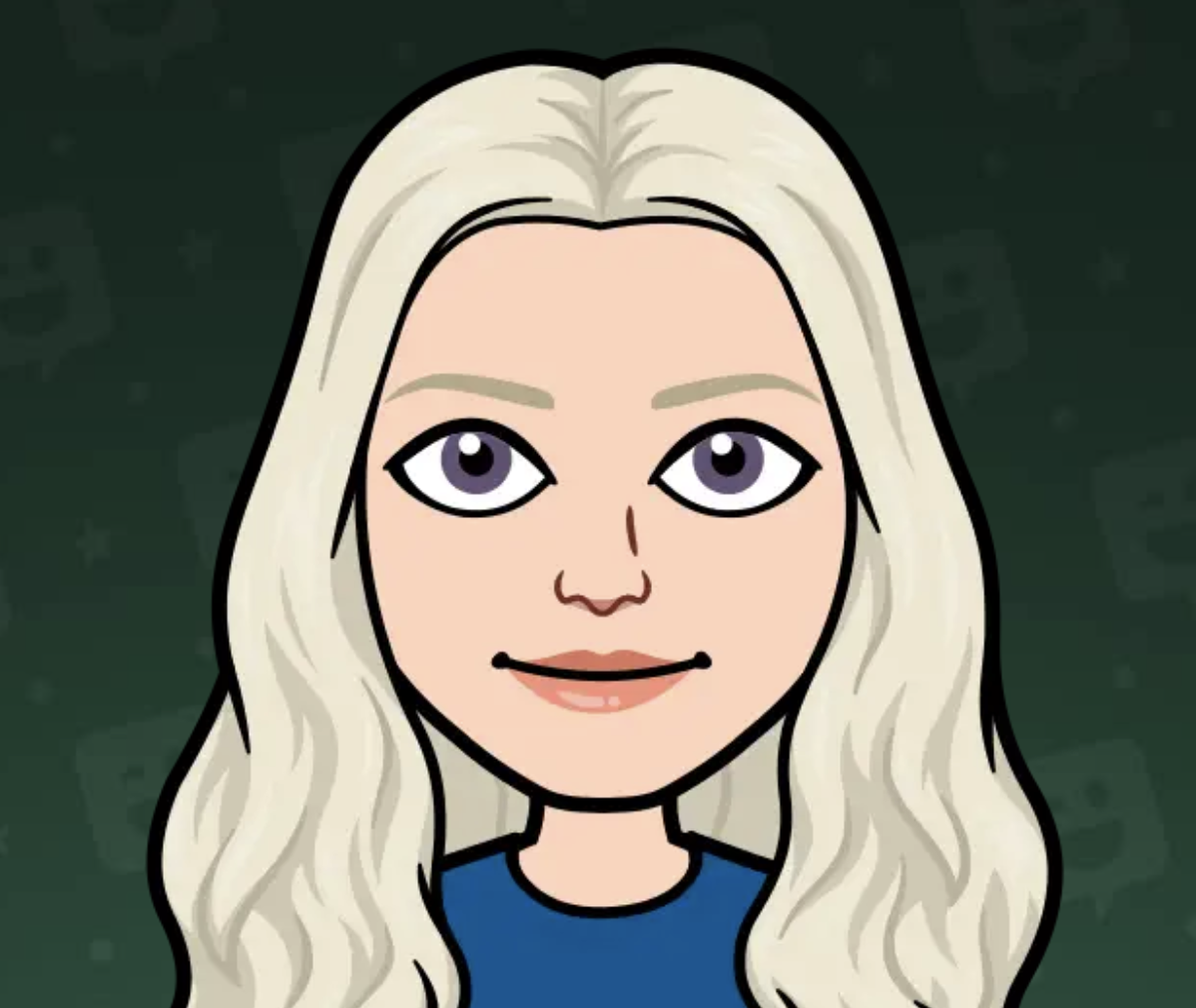}
  \caption{A Bitmoji version of P8's avatar that highlights her disability identity as a person with Albinism, reflecting her physical world appearance.}
\end{subfigure}\hfill%
\begin{subfigure}{.45\textwidth}
  \centering
  \includegraphics[height=1.8in]{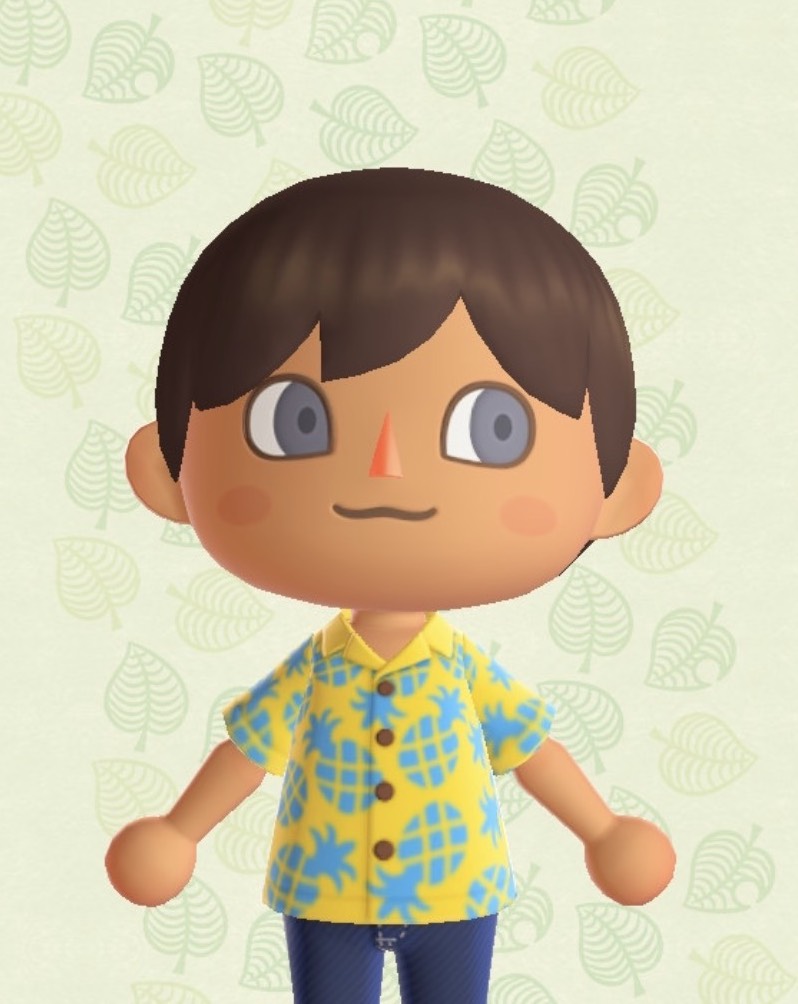}
  \caption{An \textit{Animal Crossing} version of P8's avatar that highlights her racial identity as a person who is South Asian.}
\end{subfigure}\hfill%
\caption{These two avatars both represent P8 but highlight different parts of her identity. Her eyes remain a unifying purple in both.}
    \label{fig:p8}
    \Description{Two avatars. The first is a woman with pale skin and very light blond hair and purple eyes. The second is a woman with brown skin and black hair, and the same purple eyes.}
\end{figure*}

To P4, this meant selecting between their gender or disability identity expression. Since they could only pick one identity to represent well, they chose their gender expression, explaining that their disabilities are primarily invisible and therefore a challenge to represent with current avatar platforms. Similarly, P8, a South Asian person with albinism, chose to represent her race more than her disability (see \autoref{fig:p8}):

\begin{quote}

  \textit{``I couldn't present well enough [with avatars] to look albino, so I just made myself brown, and then I think I got fond of presenting as South Asian, so then I intentionally started doing it more,''} ---P8.

\end{quote}

Because one of the prominent ways that albinism and South Asian heritage both present themselves is skin and hair pigmentation, it was difficult for P8 to show both at once. On some platforms (e.g., Bitmoji), P8 could reasonably represent her albinism, while on others (e.g., Animal Crossing), none of the options felt representative for her. 








\knm{Since platforms do not consistently support all marginalized identities well, poor representation is especially common when considering the intersectionality of people's minoritized identities: people with multiple minoritized identities often have to sacrifice showing one identity to show another more completely.} Without a conscious effort to make sure that each of these identities is represented well individually \textit{and} when combined, platforms risk further othering already minoritized groups.

\subsubsection{People Wanted to Use Avatars to Convey Fluctuating Abilities}

People with disabilities have fluctuating abilities and symptoms throughout the day. Participants discussed that they wanted this variability reflected in avatars: 
\begin{quote}
    \textit{``None of us are static ... that's not my inner world experience all day, every day. If I actually want to be able to represent my disabilities, they aren't static either...''} ---P10.
\end{quote}
Many participants wanted to use avatars to convey their abilities and access needs at a specific point in time. The symptoms and abilities that they wanted to represent included pain, mobility, fatigue/energy, focus/attention, and ability to communicate verbally. To show acute moments of things like fatigue or inattention, some participants wanted to send their avatar with phrases like \textit{``I have to go now,'' or, ``I'm no longer paying attention to you.''} (P9). 
Another commented, 
\begin{quote}
    \textit{``I would have [one version of my avatar] for when I'm nonverbal, one for when I have a migraine, one for when I'm triggered and just need to withdraw,''} ---P10.
\end{quote}

\begin{figure*}
\captionsetup{justification=centering}
\begin{subfigure}{.45\textwidth}
  \centering
  \includegraphics[height=1.8in]{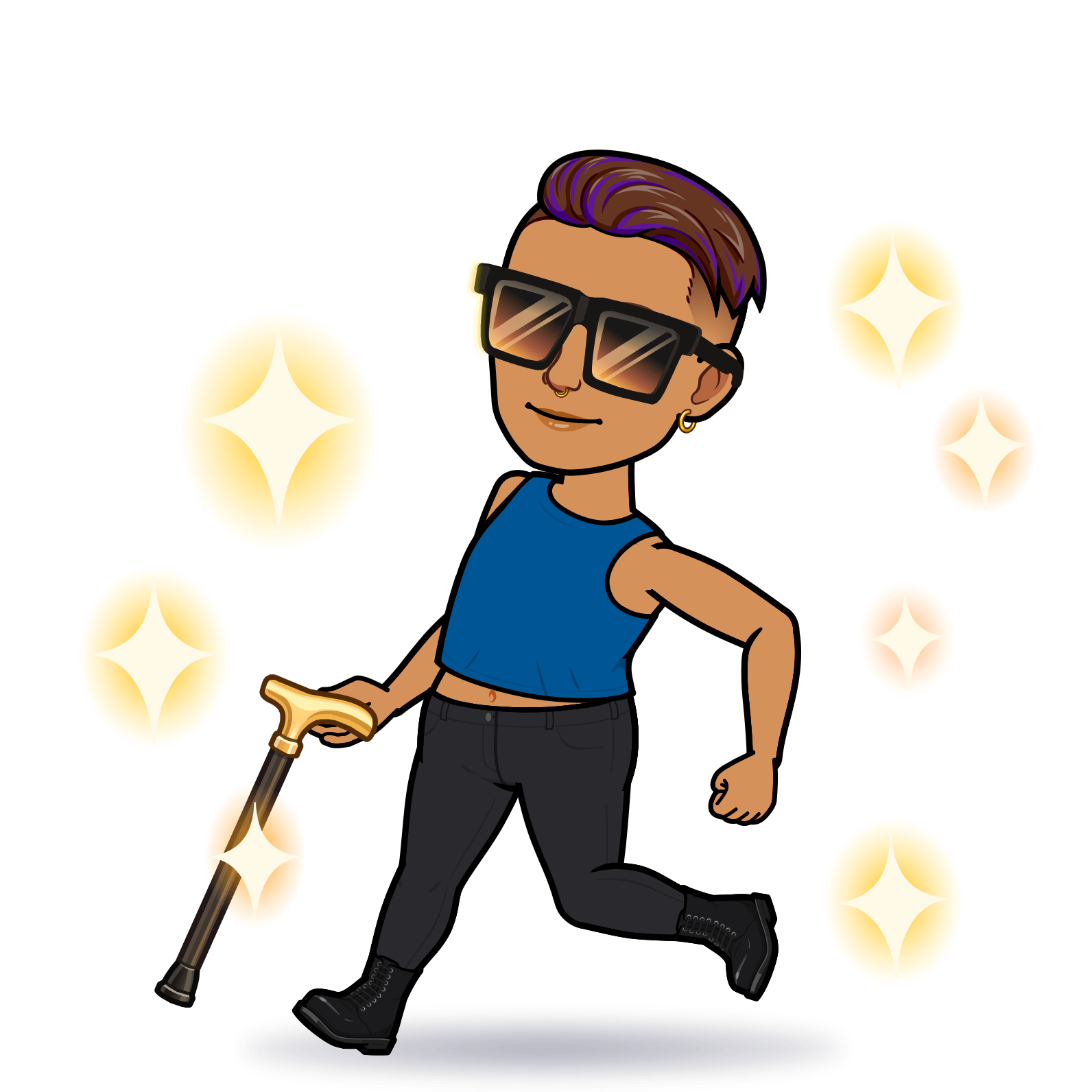}
  \caption{An avatar that signals the need to use a cane for mobility at a given time.}
\end{subfigure}\hfill%
\begin{subfigure}{.45\textwidth}
  \centering
  \includegraphics[height=1.8in]{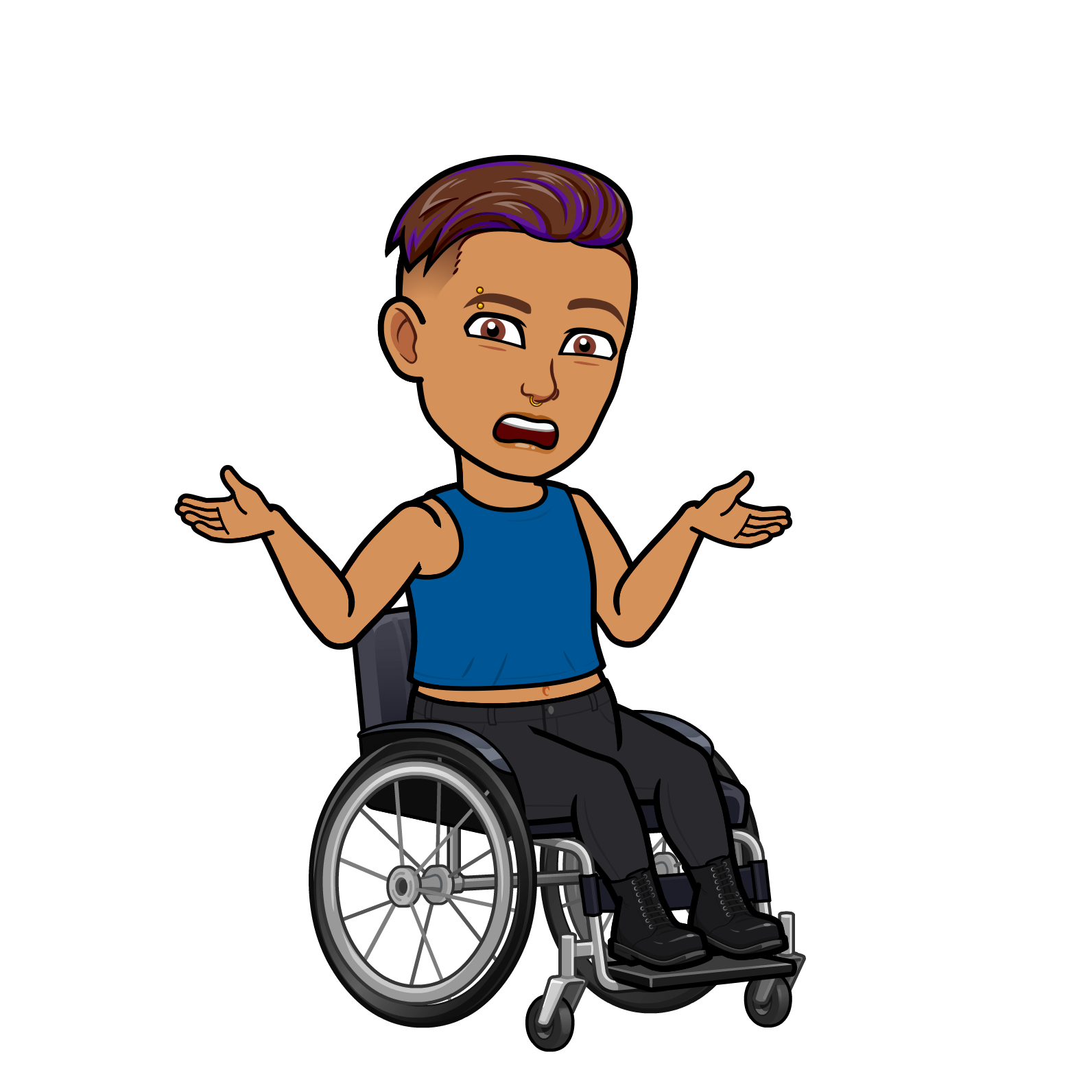}
  \caption{An avatar that signals the need to use a wheelchair for mobility at a given time.}
\end{subfigure}\hfill%
\caption{These avatars show how one might signal their mobility needs in their avatar via the assistive technology they use.}
    \label{fig:mobility}
    \Description{Two pictures of the same avatar, a queer presenting person with brown skin and hair with a blue crop top and black leggings. In the first, they are strutting with sunglasses and a cane. In the second, they have a wheelchair with an inquisitive expression on their face.}
\end{figure*}

P1 and P10 specifically were interested in using phrases like ``one spoon left'' or showing cartoon spoons. This phrase references Spoon Theory \cite{miserandino2003spoon}, which is a common way to express the inability to do things because of symptoms in the chronically ill community. Other participants, like P9, were interested in signaling their current capabilities, such as their energy level, as an ongoing ``status'' with their avatar rather than specific moments of distress. To convey status, P4 suggested adopting the traffic light system, which is common within the Autistic community, where green indicates a higher level of energy left for interacting with people and red indicates preferring no interaction. Finally, several participants with fluctuating mobility wanted to be able to convey their current ability level by showing the assistive technology they are currently using (see \autoref{fig:mobility}):
\begin{quote}
    \textit{``I could imagine indicating how mobile I was at any given time, by having the cane, or possibly using a wheelchair, or a scooter as those additional options to indicate where I am in that situation today,''} ---P9.
\end{quote}

\knm{Interestingly, the ability to show access need fluctuation could also support fluctuating gender identities, which was desired by P4, P6, P9, and P17. For example, P4 appreciated the ability to make small tweaks to their avatar (e.g., changing the broadness of shoulders or jaw shape) to best represent their gender day-to-day in a video game that allowed them to easily switch between avatars.}

Using avatars to signal current abilities/symptoms via preset stickers or persistent status indicators allowed participants an easy way to make the more hidden, internal aspects of disability visible to those they communicate with. \knm{This feature also benefits people with fluctuating gender identities in choosing how they present to other people, which is especially helpful to disabled, gender-queer people.}

\subsubsection{People Shared Avatars as a Personal, Nonverbal Communication Method}

Several participants described instances where communicating with avatars led to more accessible conversations. For example, one participant with an intellectual or developmental disability and two who are neurodiverse found that there were times when sharing visual avatars was significantly less taxing than putting together sentences (e.g., when they were nonverbal).
For P10, their dyslexia motivated their preference for avatar-based communication:
\begin{quote}
    \textit{``It was just easier to communicate ... not having to use letters and words and not having to engage that part of my brain when my dyslexic brain is exhausted ... Also, I cannot tell you how many times I still get people correcting my grammar ... They're just shaming [me]. To not have to use letters to convey a message is brilliant,''} ---P10.
\end{quote}
P10 was able to avoid situations that often led to judgment or stigma related to their disability by sending avatars with preset phrases rather than typing out messages. In both cases, sending avatars with preset messages was more accessible.

Another deaf participant found communicating orally to be challenging at times, so he started using his avatars in work contexts: 
\begin{quote}
    \textit{``I use it a lot in relating with my customers ... in the chat. I would just send them an avatar to show. I feel it's more [expressive] since I really can't hear. The avatar helps paint my mood,''} ---P13.
\end{quote}
In text mediums, nonverbal signals like tone and facial expressions are easily lost \cite{kiesler1984social}. Avatars allowed P13 to be more personal while remaining in an accessible text medium.


Regardless of whether their disability was represented in their avatar, participants with various disabilities found that the medium of avatars, due to their pictorial nature, could be more accessible than alternatives.

\subsection{Portraying Disability in Avatars}

Participants were usually very intentional in creating their avatars, whether that meant highlighting their disabilities or obscuring them. \knm{In cases where participants were able to focus on self-expression over impression management, they described the ideal ways they would like to share their disabilities.} We describe how participants use current features of avatar platforms to show disability identities and how they would like avatar platforms to better support their identity expression.

\subsubsection{Current Use: Showing Assistive Technology and the Full Body}
Many participants showed their disabilities in avatars through their assistive technologies and how they posed their bodies. Assistive technologies can become an intimate part of their identity, especially when they are prominent in their life: 
\begin{quote}
    \textit{``When people think of me, they think of my wheelchair, my scooter, me, and my service dog [name], because she goes everywhere I go,''} ---P5.
\end{quote}
With current avatar platforms, this type of identity expression is most often only available for hearing aids or mobility aids, and several participants wanted more variety in the assistive technologies supported. The other main strategy that helped participants feel represented was posing their avatar in ways that showed their whole bodies, rather than just above their shoulders, though this is not a common option in many avatar platforms \cite{zhang2022s}. Showing the full body allowed access tools like compression socks (P17) or a wheelchair (P6), or a person's fatness to become visible. In general, participants were interested in selecting poses for their avatars that communicated their disability.

However, outside of these two cases, participants found that platforms did not support the variety of ways they could imagine representing their disabilities in avatars. Their ideas mainly fell into four different dimensions of avatars: (1) body modifications, (2) assistive technologies, (3) symbols, and (4) backgrounds and actions.

\subsubsection{Desired Use: Modifying Parts of the Avatar Body to Show Disability}

Users can often customize the physical characteristics of the body, including body shape/type, skin tone, facial features, hairstyle, and hair color. Participants expressed interest in being able to customize avatar body proportions to better represent their disability identities. For example, the ability to customize limb length would allow better representation of people with dwarfism, certain muscular conditions, and amputees. In the same vein, participants who identified as fat commented that, when adjusting fatness is an option, it is often poorly implemented:
\begin{quote}
    \textit{``The ones that [show fatness] are typically terrible ... I think one of the really beautiful things about fatness is that it's extremely diverse. Fat does not sit on two people's bodies the same way ... it would just be amazing to have something where you could actually change the proportions of the body,''} ---P3.
\end{quote} 
P3's comment suggests that customizing the amount of fat on different body parts would allow her to feel better represented. While this customization is highly important for people who consider fatness and disability core part(s) of their identity, it also better represents any person who aims to best match their physical appearance and whose body type does not  conform to the most commonly represented thin body with four limbs.  

Further, two participants commented that exact skin tone was critical to representing their disabled selves. P18 sees their pale complexion as an important indicator of their disability, since their chronic illness symptoms require them to stay inside frequently. Additionally, they wanted more fine-grained customization around being able to show dark skin under their eyes, a consequence of their chronic illness. Similarly, P8 has albinism and her pale skin is a key representation of her identity. However, it can be difficult to find the correct skin tone: she often could not find her pale skin tone rather than one that is ``paper-white.''

\subsubsection{Desired Use: Representing a Broad Variety of Customizable Assistive Technologies}

Our participants often had the goal of matching their actual physical appearance as much as possible, and without the ability to represent their assistive technology, a core part of their appearance is omitted in avatars. However, participants cautioned that, to be done well, key aspects of virtual assistive technologies must match their real-life counterparts. For example, some people customize their assistive technologies in real life as a fun form of expression; P5 had a collection of colorful canes, and P12 described people selecting colorful ear molds for hearing aids. \knm{While most participants did not use platforms that support this customization, P6 appreciated a platform that allowed him to customize the appearance of the wheelchair to be rainbow-colored such that he could show both his disability and queer pride.} On the other hand,
P5 and P6 were put off when assistive technologies in virtual spaces were incorrectly designed: 
\begin{quote}
    \textit{``Whenever somebody is posed in a [wheel]chair and [the wheelchair] doesn't fit them, it just does not look right at all and that is so distracting. It's like, what if an avatar had on clothes that were too short or too big?''} ---P5.
\end{quote}
Details like color and the fit of the technology to the person matter. P6 further highlighted that, in order to properly represent assistive technology, designers should work with disabled people and communities in the design process.

\subsubsection{Desired Use: Adding Symbols to Share Disabled Experiences}

Adding symbols to the avatar, especially words and objects, is a powerful tool that participants wanted to use to convey their disabled experiences and share disability culture (see \autoref{fig:annotations}). For example, P10 wanted to show words in Bionic, a dyslexia-friendly font. P9 wanted to share that they had ADHD and were interested in leaning into existing stereotypes to share common phrases that demonstrate their experience, like looking away from a person and saying ``squirrel.''\footnote{Note that while some people embrace this stereotype, others find it harmful or offensive \cite{hey2019is}.}

Symbols were particularly helpful for people with invisible disabilities and symptoms. P15 wanted to be able to add a symbol or flag\footnote{The \href{https://www.respectability.org/2022/07/disability-pride-flag/}{disability pride flag}, to our knowledge, is not represented in any of the most used platforms by our participants} as an annotation to the side of the avatar to represent his invisible disability, similar to queer pride symbols. P10 wanted the ability to put little pins and needles poking into all parts of their body at once to signify how it feels to have certain types of chronic pain. P10 also saw an opportunity to play with the cartoon avatar's appearance to better represent how people are feeling on the inside: 
\begin{quote}
    \textit{``People, once they are pregnant, their feet grow and other parts of their body don't work the same way that they used to. Being able to represent that your feet are absolutely just destroyed today, by enlarging them and making them red is a way of helping people decide for themselves what visibility looks like for their disability.''} ---P10.
\end{quote} 
P1 suggested using word annotations to incorporate disability culture into avatars in existing poses. She explained that most of the ``tired'' stickers reflect how she feels reasonably well, but that she would use those stickers more if they had a phrase like ``no spoons.'' 
Symbols allowed participants to add bits of their disability or culture to their avatars.

\subsubsection{Desired Use: Selecting Actions and Backgrounds that Represent Disability Culture or Experience}

Participants also wanted to use an avatar's action and location to convey their disabled experience. For example, P7, P12, and P13, all of whom are d/Deaf, wanted avatar platforms to support short signed phrases in American Sign Language (ASL) as preset animations, or even allow the user to animate their avatar by recording themselves signing longer thoughts. For another participant, actions could allow them to represent their neurodiversity by showing a person ``flapping,'' which is a common stimming\footnote{\href{https://www.verywellhealth.com/what-is-stimming-in-autism-260034}{Stimming} is repetitive behaviors done as a form of self-stimulation or self-soothing.} behavior for some neurodiverse people (P4). Finally, people commented that avatars' backgrounds can provide further context into a disabled person's experience. P1 wanted her avatar to be shown in a bedroom because she is a self-described ``sleepy person'' in part because of how her disabilities manifest in her life, and P9 wanted to be shown with a slightly cluttered environment with whiteboards everywhere since they have ADHD and utilize distributed whiteboards to remember things. These examples show the creative potential that the avatar's action and location have to represent disability for a diverse range of people.

\subsection{Avatar Platforms and Physical World Social Norms Shape Each Other}

Without care, avatar platforms can easily encode the biases of the physical world, leading minoritized users to purposefully craft their avatars in ways that counteract stereotypes. However, we also saw cases where participants' positive representation in avatars led to positive responses in the physical world.

\subsubsection{Options and Defaults in Avatar Platforms can Perpetuate Physical World Biases}

Avatar systems take on and create platform norms that are informed by broader social context. Platforms determine who does and does not get to exist in the virtual space through the options they provide for people to represent themselves. Consequently, avatar platforms can replicate social biases through the options they surface to users. 
Determining which features of an avatar are editable and which are not limits who can be represented in a system. P3 explains: \begin{quote}
    \textit{``A lot of avatar makers, when they're representing body size they'll like, `go for height,' they're fine with height diversity but then they'll give you a very narrow range for fatness. I often feel it's like the subtext is who would pick to be fat? It's like, `When you're not giving me the option to do that, you're reinforcing that I shouldn't choose it.'''} ---P3.
\end{quote}
In this case, the omission of fatness as an option not only frustrates P3, as she cannot fully represent how she appears in the physical world, but also perpetuates biases in broader society around fatness as undesirable \cite{kyrola2021fat}.

Relatedly, having some options as ``defaults'' versus add-ons affects who feels like they belong on a platform. Two participants explained that there were better body or hair options to represent themselves and their identity, but they had to be purchased or unlocked. For example, platinum blonde hairstyles (which are closer to P8's hair color) were costly upgrades from the ``standard'' hair colors in the game \textit{Animal Crossing}. In addition to leaving players frustrated with an unrepresentative avatar, this experience actually deterred P8 from attempting to make a visually realistic representation of herself. Locking certain customization features behind paywalls can thus imply who can exist easily or ``by default,'' and who must incur more friction (e.g., monetary cost, time, effort) to exist in a virtual world.

Finally, blatantly harmful stigmas from the physical world can be built into platforms. For example, P10 commented on how mental illness is represented in the video game series \textit{The Sims}. Sims avatars can have personality characteristics, and the only one that comes close to mental illness is ``insane.'' P10 explains their frustration:
\begin{quote}
    \textit{``How that is represented in that game is somebody who is angry, erratic, emotional. It's basically every negative trope about having a mental health diagnosis or disability. It's [the] only ... [disability] representation that you get and it's that word,''} ---P10.
\end{quote}
While disability representation matters in avatars and games, this example demonstrates that something is not always better than nothing: careless representations of disability in avatars propagate harmful stereotypes.

\subsubsection{Stereotypes and Stigma in the Physical World Affect Avatar Expression}

Since avatars have the ability to perpetuate stereotypes, participants were conscious of avoiding stereotypes in their own representations. 
P6 expressed his frustration with stock photos of wheelchair users perpetuating stereotypes about their abilities: 
\begin{quote}
    \textit{``A lot of times it's like someone sitting still, or even someone being pushed by someone else. [I want] options to emphasize like, ``Yes, this is a person who can, in fact, move in a wheelchair,'' ''} ---P6.
\end{quote}
Similarly, both of our blind participants and one of our participants who uses a wheelchair wanted to show their avatars doing specific activities to combat the stereotype that disabled people sit around or cannot have active lifestyles. P8 would sometimes intentionally change her skin tone to have more pigmentation than she has in real life to match her South Asian heritage: 
\begin{quote}
    \textit{``I think it's because I don't get the chance to make first impressions in real life ... They look at my last name ... and I have to explain it, and I have to sometimes prove [my ethnicity] to them for some reason. I'll speak my mother's tongue ... If I'm playing with strangers, and they don't know that I have albinism, I can just pretend to look like who I am on the inside, and they wouldn't know. I still get to have that experience where people, by default, if they look at my little character, they're like, `Okay it's a brown girl.' I'm changing their stereotypes of me on purpose,''} ---P8.
\end{quote}
In this example, P8 utilized the ability to adjust her skin tone to be different than it is in reality, to  reflect an authentic part of herself that isn't always visible (i.e., her South Asian heritage). Through this process and the level of appearance customization that avatars support, she can control the stereotypes that she might encounter when interacting with others.

\subsubsection{Avatars can Shape Physical World Norms and Feelings about Disability Identity}

At the same time, avatars have the ability to influence the physical world. Participants commented on ways that their disabilities being represented virtually led to physical world gains. For example, virtual representation emboldened some participants to feel and/or express more pride in their disability. P3 spoke about how she shows stigmatized elements of her identity in avatars as an act of reclamation:
\begin{quote}
    \textit{``It's nice to be able to incorporate things about your body that have been stigmatized that you have become more comfortable with as a way to take away the stigma,''} ---P3.
\end{quote}
In one example, P18 felt stigma around his hearing aids and chose not to include them in his avatar. However, once hearing aids in avatars were adopted broadly, this reduced the stigma to the point that P18 added hearing aids to his avatar. Some participants represented their disabilities in their avatars to combat stigmas and eliminate them in both virtual and real spaces.

In each of these examples, we see that representation in avatars can have real, positive effects on the user who wants to feel better represented.

\subsubsection{Excluding People is Harmful to Users and Platforms}

When platforms do not actively choose inclusive avatar designs, the resulting exclusion has negative effects for the user and the platform. For example, P3 explained how disengaged she felt when playing fitness games that allowed for only thin avatars: \textit{``My body is the one doing the work here. It is actually a fat person doing this.''} Some participants commented that they could not use noninclusive technologies, especially on days they already feel sensitive: 
\begin{quote}
    \textit{``I definitely think that the avatars creation is not something I would do on like [a] `Bad Body Day'. If I am hearing the siren song of fitness one day, that's not a time to go and make an avatar. It's a recipe for feeling shitty about myself,''} ---P3.
\end{quote}
This example shows that avatars that do not look like users, but rather what the norms in our society say they should ``want'' to look like, can negatively impact people's body image or mental health. Other participants fully eschewed technologies that were inaccessible (e.g., P0 does not use avatars much because he uses a screen reader and so few avatars have alt text) or non-representative (e.g., P9 refuses to make a Meta avatar because they cannot be shown as fat enough). These examples demonstrate that poor representation in avatars does not just have negative effects for the user; it actively loses platforms users.

\section{Discussion}

\knm{Our data provides insights into how people with disabilities want to use avatars and represent their disability and other important identities. Our participants largely spoke of using avatars in social applications where, similar to findings from prior work \cite{zhang2022s, freeman2021body}, they aimed to make realistic, or slightly more idealized avatars \cite{turkle1994constructions, neustaedter2009presenting}. Aligned with research on gender and racial stereotypes in avatars \cite{mcarthur2014everyone, consalvo2003s, mcarthur2015avatar}, their experiences demonstrate that they often encounter stereotypes about their disability when crafting their avatars and have to make very conscious decisions to avoid them. Our participants frequently chose to disclose their disability in their avatar to show representation to others, show pride, and authentically express themselves \cite{zhang2022s}. Moving beyond prior work, we found that participants wanted to harness avatars to communicate about their disabled experience and to communicate in more accessible ways than they could otherwise. Specifically, while prior work focuses on BLV and DHH participants \cite{zhang2022s}, our work with a considerable number of neurodiverse, chronically ill, and mentally ill participants provides insights into how a broader set of disabled people want to be represented, especially those with invisible disabilities. We now discuss how avatars can best represent disabled and intersectional identities, how to support signaling group identity membership in avatars, and design recommendations for avatar platform makers.}

\subsection{Representing Intersectional, Fluctuating Identities}

\knm{Audre Lorde and Sins Invalid emphasize the importance of the fact that disabled people ``do not live single issue lives,'' but rather ones with complex intersecting facets of identity \cite{lorde1982learning, Sins_Invalid_2019}. Building on research that discusses the importance of examining multiple identities \cite{bennett2021s, abebe2022anti} and dimensions of identity individually \cite{mcarthur2015avatar, zhang2022s}, we present the unique issues that arise for these people in avatar representation. Specifically, findings from our sample of almost exclusively multiply minoritized people emphasize the importance of understanding intersectional issues, especially in the space of identity representation. Avatar platforms that did not support minoritized people well-created issues around what identities to show, forcing participants to choose between their identities, or what platforms to use. These issues led participants to avoid platforms on which they could not show certain identities.}

\knm{However, if platforms did support diverse representation, avatars were a useful tool for participants that could act as filters: morphing appearances to occlude some identities while bringing others to the forefront. We highlight three features of avatars that are key in supporting disabled people's desired identity presentation.}

\knm{First, avatars can be \textbf{hyper-customizable}, which allows people to be intentional in how they craft their appearance to show their identities. We extend prior work on customization options for race and gender~\cite{morgan2020role,  mcarthur2015avatar, mcarthur2014everyone}, and disability disclosure for BLV and DHH people in VR~\cite{zhang2022s}, by demonstrating how avatars could represent invisible disabilities and elucidate access needs. 
This can cross many different characteristics of avatars, from skin tone and bruising, to poses, to assistive technology use---with some individual features being able to support multiple identities at once (e.g., a rainbow wheelchair showing both disability identity and queer pride). }

\knm{Second, avatars' ability to \textbf{deviate from reality} allowed people to better represent or hide aspects of their identity. While prior work discusses switching identities to control how others perceive them \cite{dimicco2007identity, turkle1999cyberspace, bruckman1996gender, haimson2016constructing, edwards2020three, zhang2022s}, our work highlights how these choices might not always be unconstrained; some identities can be difficult to express at the same time (e.g., P8 drastically altering skin tone to switch between her race and disability identities), and others are not supported by avatar platforms. With the public's oftentimes stigmatized view of disability, people with disabilities have to be particularly selective in whether and when to show an identity. Sometimes, participants wanted to express pride in their disability identity (e.g., demonstrating disability representation to younger colleagues) but wanted to obscure it completely when they felt unsafe (e.g., hiding disability in online forums). As such, participants needed to balance which identities to present in their avatars, ``editing oneself''~\cite{bullingham2013presentation} depending on goals of authentically expressing themselves and/or influencing the impression they want to give off in certain situations.}


\knm{Finally, the ability to \textbf{easily update} avatars is critical for bringing different identities to the forefront or background depending on how they want others to perceive them. This was a critical feature for sharing nuances of disability like a person's current abilities or access needs (e.g., showing mobility based on the avatar's pose or assistive technology). Indeed, each of these three features can support both disability identity \textit{and} presenting multiple, intersecting identities.}

\subsubsection{Future Considerations}

Many of the ways people want to use avatars to better represent identity rely on the ability to easily make changes that persist for different lengths of time. Some changes might span days or months, while others might be only needed for one message. For example, someone who wants to send a message to a friend showing how tired they are might only make a small change like adding the words ``no spoons left'' or a red light to their avatar. They might not want this small annotation to their avatar to persist after this message is sent, since it represents something that changes frequently. More long-lasting or significant changes (e.g., changing gender expression or dying hair) might require updating the avatar at its base rather than using annotations. Enabling users to modify their avatars at different time scales---perhaps through saved configurations that users can swipe between---is key to making it easier for them to express their identities. \knm{Future research can consider how to make the process of creating and selecting the most applicable, highly-customized avatars as easy as possible (e.g., can we use sensing differences in ability, like gait, to predict what kind of assistive technology the user will want the avatar to use, like a cane).}

\subsection{Intentional Community Signaling in Avatars}

People were very intentional in how they presented or wanted to present their identities to signal belonging to a certain community. To do so, they sometimes referenced community cultures. Queer people incorporated the established flag colors into their avatars, and chronically ill participants wanted to reference ``Spoon Theory'' by adding cartoon representations of spoons \cite{miserandino2003spoon}. In addition to signaling identity with one community, participants also were conscious about not propagating stereotypes that they found harmful about their identity. For example, blind participants actively tried to show themselves being active to combat inaccurate stereotypes about their lives. 

To support community-based identity expression, platforms need to consider how to facilitate in-group signaling while avoiding problematic or offensive representations. This problem is nontrivial given that what people find representative of their identity varies and conflicts between groups; even in our sample, a stereotype for one group was the ideal representation for another. For instance, while blind participants wanted more active avatars to combat stereotypes, people who were chronically ill were frustrated by the large number of active avatars, and wanted more options for low-intensity activities and poses (e.g., lying down, reading a book). 
The same issue can arise within a specific community; a mode of signaling might be fun for one person but offensive to another. Though P9 wanted a playful representation of themselves saying ``squirrel,'' which is commonly associated with ``ADHD,'' others within the ADHD community find that stereotype offensive and incorrect \cite{hey2019is}. 

To address these issues, platforms should consider when it is appropriate to pre-label something as ``an option for people of this identity'' (e.g., chronically ill content). Content created and branded as supporting people of a specific identity can be a powerful way to help users feel seen. It can also be an access benefit because it saves the user the time and effort of finding or making the content \cite{zhang2022s}. At the same time, labeling content as being ``for a certain group of people'' risks offending people who disagree with that label. Certain strategies could mitigate these negative aspects of creating curated identity-based content. First, platforms should consider surveying a broad group of people within the corresponding community about potentially sensitive content. Disabled communities hold the expertise to identify desirable content versus potentially harmful or offensive. Second, platforms can allow people to make their own presets or curated content. While this option does not solve the challenges that minoritized people might face in their initial search for content, people can still benefit from fast, easy access to the identity-expressing and affirming content they find.

\subsection{Recommendations for Avatar Platforms}
From these findings, we present a list of considerations for avatar platforms around disability representation:
\begin{enumerate}

   \item \textbf{Support customizability and easy change}
   \begin{enumerate}
      \item \textbf{Allow customization of body part sizes:} notably, allowing users to change the way weight sits on the body, and the length/presence of limbs is important for disabled and fat representation.
      \item \textbf{Provide default poses that show more than the head:} many aspects of disability, such as assistive technology use and body size, are most visible from the neck down.
      \item \textbf{Allow color pickers wherever possible:} participants felt limited and frustrated in selecting their skin and hair color from a limited palette. Sometimes colors of other elements, like assistive technology, were important for people's self-expression.
      \item \textbf{Allow users to make multiple, saved versions of avatars:} participants wanted different versions of avatars that expressed different states of disability (e.g., symptoms, assistive technology use), or may or may not present disability. Switching between versions also must be easy to allow for seamless identity representation in different contexts.
   \end{enumerate}

   \item \textbf{Make disabled and multiply minoritized users feel welcomed}
   \begin{enumerate}
      \item \textbf{Do not block identity representation with paywalls:} do not require users to pay for features that are important for representing a core identity (e.g., hairstyles, skin tones, assistive devices)
      \item \textbf{Make avatars accessible to create and consume:} this includes making sure avatars have alt text, the associated text has proper color contrast, and that avatars can be created and consumed with only keyboard access.
      \item \textbf{Ensure features can be combined seamlessly:} for example, check that people with a variety of body types still look realistic when posed in a wheelchair. Check that assistive devices look natural on people with a variety of skin tones. This is especially important for multiple minoritized avatar users.
   \end{enumerate}

   \item \textbf{Engage with disabled communities}
   \begin{enumerate}
      \item \textbf{Work with disabled people:} disabled communities can help with concretizing designs for showing disability in avatars (e.g., assistive technology).
      \item \textbf{Avoid perpetuating harmful stereotypical representations:} talking with disabled communities can help avoid these pitfalls.
   \end{enumerate}

\end{enumerate}

Designing avatar platforms such that they are inclusive of all types of identity is a challenging issue. Our design recommendations are starting guidelines, and should be considered in tandem with feedback from disabled communities. We encourage designers to consider how avatar platforms can maintain or enhance the key functionality of avatars in supporting intersectional, fluctuating identities and community signaling.

\section{Limitations}

While our interviews and findings revealed deep, rich examples of people's experiences and perceptions of disability representation in avatars, our work also has some limitations. For example, our sample was self-selected, and thus skews towards people who are younger and people with strong disability identities and  pride. This is an important group to represent, as they likely have strong opinions on how they want their identities represented. However, future work could build on our results with broader-scale surveys of more people with various perceptions of their disability identity, while also allowing for greater disability and geographic diversity. Surveying a broader sample is an important follow up, as our sample of 18 participants is not fully representative of any one disabled community, as is participatory work to understand the specific execution of our guidelines on platforms. Further, our discussions with interviewees mainly focused on avatars in non-mixed reality social media/social app contexts. Some of our insights likely extend to mixed reality contexts and video games, though they may require more specific needs and considerations for the avatars they provide.

\section{Conclusion}
Because avatars are representations of ourselves in virtual spaces, it is critical that core parts of identity, like disability, can be shown in this medium. We interviewed 18 people with disabilities or related identities about their preferences in representing their disability identities and using avatars. Participants described creative ways they can use many dimensions of avatars to convey their disability status and experience, including body modifications, poses, symbols, and actions and backgrounds. Overall, participants wanted to share their disability identities and experiences, oftentimes using avatars to signal their abilities, symptoms, or disability pride. However, participants had to feel safe to do so and supported by platforms in expressing their identities. If participants had multiple minoritized identities, avatars uniquely enabled them to choose which to highlight to others, but also required trade-offs since most platforms did not support all of their identities well. Finally, we discussed how avatars' flexibility, ease of use, and ability to bend reality were helpful in representing disability, especially for people with fluctuating abilities and identities. We summarize our insights into design recommendations for future researchers and designers to build avatar platforms with disability (and other minoritized) representation in mind.

\begin{acks}
We would like to thank all of our participants and Maarten Bos for their time and insights.
\end{acks}

\bibliographystyle{ACM-Reference-Format}
\bibliography{references}


\begin{thebibliography}{55}


\ifx \showCODEN    \undefined \def \showCODEN     #1{\unskip}     \fi
\ifx \showDOI      \undefined \def \showDOI       #1{#1}\fi
\ifx \showISBNx    \undefined \def \showISBNx     #1{\unskip}     \fi
\ifx \showISBNxiii \undefined \def \showISBNxiii  #1{\unskip}     \fi
\ifx \showISSN     \undefined \def \showISSN      #1{\unskip}     \fi
\ifx \showLCCN     \undefined \def \showLCCN      #1{\unskip}     \fi
\ifx \shownote     \undefined \def \shownote      #1{#1}          \fi
\ifx \showarticletitle \undefined \def \showarticletitle #1{#1}   \fi
\ifx \showURL      \undefined \def \showURL       {\relax}        \fi
\providecommand\bibfield[2]{#2}
\providecommand\bibinfo[2]{#2}
\providecommand\natexlab[1]{#1}
\providecommand\showeprint[2][]{arXiv:#2}

\bibitem[Abebe et~al\mbox{.}(2022)]%
        {abebe2022anti}
\bibfield{author}{\bibinfo{person}{Veronica Abebe}, \bibinfo{person}{Gagik
  Amaryan}, \bibinfo{person}{Marina Beshai}, \bibinfo{person}{Ali~Ekin Gurgen},
  \bibinfo{person}{Wendy Ho}, \bibinfo{person}{Naaji~R Hylton},
  \bibinfo{person}{Daniel Kim}, \bibinfo{person}{Christy Lee},
  \bibinfo{person}{Carina Lewandowski}, \bibinfo{person}{Katherine~T Miller},
  {et~al\mbox{.}}} \bibinfo{year}{2022}\natexlab{}.
\newblock \showarticletitle{Anti-Racist HCI: notes on an emerging critical
  technical practice}. In \bibinfo{booktitle}{\emph{CHI Conference on Human
  Factors in Computing Systems Extended Abstracts}}. \bibinfo{pages}{1--12}.
\newblock


\bibitem[Arts(2014)]%
        {sims4}
\bibfield{author}{\bibinfo{person}{Electronic Arts}.}
  \bibinfo{year}{2014}\natexlab{}.
\newblock \bibinfo{title}{The Sims 4}.
\newblock
\newblock


\bibitem[Bailey et~al\mbox{.}(2009)]%
        {bailey2009avatar}
\bibfield{author}{\bibinfo{person}{Rachel Bailey}, \bibinfo{person}{Kevin
  Wise}, {and} \bibinfo{person}{Paul Bolls}.} \bibinfo{year}{2009}\natexlab{}.
\newblock \showarticletitle{How avatar customizability affects children's
  arousal and subjective presence during junk food--sponsored online video
  games}.
\newblock \bibinfo{journal}{\emph{CyberPsychology \& Behavior}}
  \bibinfo{volume}{12}, \bibinfo{number}{3} (\bibinfo{year}{2009}),
  \bibinfo{pages}{277--283}.
\newblock


\bibitem[Barnes(2019)]%
        {barnes2019understanding}
\bibfield{author}{\bibinfo{person}{Colin Barnes}.}
  \bibinfo{year}{2019}\natexlab{}.
\newblock \showarticletitle{Understanding the social model of disability: Past,
  present and future}.
\newblock In \bibinfo{booktitle}{\emph{Routledge handbook of disability
  studies}}. \bibinfo{publisher}{Routledge}, \bibinfo{pages}{14--31}.
\newblock


\bibitem[Bennett et~al\mbox{.}(2021)]%
        {bennett2021s}
\bibfield{author}{\bibinfo{person}{Cynthia~L Bennett}, \bibinfo{person}{Cole
  Gleason}, \bibinfo{person}{Morgan~Klaus Scheuerman},
  \bibinfo{person}{Jeffrey~P Bigham}, \bibinfo{person}{Anhong Guo}, {and}
  \bibinfo{person}{Alexandra To}.} \bibinfo{year}{2021}\natexlab{}.
\newblock \showarticletitle{“It’s Complicated”: Negotiating Accessibility
  and (Mis) Representation in Image Descriptions of Race, Gender, and
  Disability}. In \bibinfo{booktitle}{\emph{Proceedings of the 2021 CHI
  Conference on Human Factors in Computing Systems}}. \bibinfo{pages}{1--19}.
\newblock


\bibitem[Braun and Clarke(2006)]%
        {braun2006using}
\bibfield{author}{\bibinfo{person}{Virginia Braun} {and}
  \bibinfo{person}{Victoria Clarke}.} \bibinfo{year}{2006}\natexlab{}.
\newblock \showarticletitle{Using thematic analysis in psychology}.
\newblock \bibinfo{journal}{\emph{Qualitative research in psychology}}
  \bibinfo{volume}{3}, \bibinfo{number}{2} (\bibinfo{year}{2006}),
  \bibinfo{pages}{77--101}.
\newblock


\bibitem[Braun and Clarke(2019)]%
        {braun2019reflecting}
\bibfield{author}{\bibinfo{person}{Virginia Braun} {and}
  \bibinfo{person}{Victoria Clarke}.} \bibinfo{year}{2019}\natexlab{}.
\newblock \showarticletitle{Reflecting on reflexive thematic analysis}.
\newblock \bibinfo{journal}{\emph{Qualitative Research in Sport, Exercise and
  Health}} \bibinfo{volume}{11}, \bibinfo{number}{4} (\bibinfo{year}{2019}),
  \bibinfo{pages}{589--597}.
\newblock


\bibitem[Bruckman(1996)]%
        {bruckman1996gender}
\bibfield{author}{\bibinfo{person}{Amy Bruckman}.}
  \bibinfo{year}{1996}\natexlab{}.
\newblock \showarticletitle{Gender swapping on the Internet}.
\newblock \bibinfo{journal}{\emph{High noon on the electronic frontier:
  Conceptual issues in cyberspace}} (\bibinfo{year}{1996}),
  \bibinfo{pages}{317--326}.
\newblock


\bibitem[Bullingham and Vasconcelos(2013)]%
        {bullingham2013presentation}
\bibfield{author}{\bibinfo{person}{Liam Bullingham} {and}
  \bibinfo{person}{Ana~C Vasconcelos}.} \bibinfo{year}{2013}\natexlab{}.
\newblock \showarticletitle{‘The presentation of self in the online world’:
  Goffman and the study of online identities}.
\newblock \bibinfo{journal}{\emph{Journal of information science}}
  \bibinfo{volume}{39}, \bibinfo{number}{1} (\bibinfo{year}{2013}),
  \bibinfo{pages}{101--112}.
\newblock


\bibitem[Chapman(2020)]%
        {chapman2020neurodiversity}
\bibfield{author}{\bibinfo{person}{Robert Chapman}.}
  \bibinfo{year}{2020}\natexlab{}.
\newblock \showarticletitle{Neurodiversity, disability, wellbeing}.
\newblock In \bibinfo{booktitle}{\emph{Neurodiversity Studies}}.
  \bibinfo{publisher}{Routledge}, \bibinfo{pages}{57--72}.
\newblock


\bibitem[Cobley(2018)]%
        {cobley2018understanding}
\bibfield{author}{\bibinfo{person}{David Cobley}.}
  \bibinfo{year}{2018}\natexlab{}.
\newblock \showarticletitle{Understanding, defining and measuring disability}.
\newblock In \bibinfo{booktitle}{\emph{Disability and International
  Development}}. \bibinfo{publisher}{Routledge}, \bibinfo{pages}{8--28}.
\newblock


\bibitem[Consalvo(2003)]%
        {consalvo2003s}
\bibfield{author}{\bibinfo{person}{Mia Consalvo}.}
  \bibinfo{year}{2003}\natexlab{}.
\newblock \bibinfo{booktitle}{\emph{It's a queer world after all: Studying The
  Sims and sexuality}}.
\newblock \bibinfo{publisher}{Glaad}.
\newblock


\bibitem[Crenshaw(1990)]%
        {crenshaw1990mapping}
\bibfield{author}{\bibinfo{person}{Kimberle Crenshaw}.}
  \bibinfo{year}{1990}\natexlab{}.
\newblock \showarticletitle{Mapping the margins: Intersectionality, identity
  politics, and violence against women of color}.
\newblock \bibinfo{journal}{\emph{Stan. L. Rev.}}  \bibinfo{volume}{43}
  (\bibinfo{year}{1990}), \bibinfo{pages}{1241}.
\newblock


\bibitem[Cullen et~al\mbox{.}(2018)]%
        {cullen2018better}
\bibfield{author}{\bibinfo{person}{Amanda Cullen}, \bibinfo{person}{Kathryn
  Ringland}, {and} \bibinfo{person}{Christine Wolf}.}
  \bibinfo{year}{2018}\natexlab{}.
\newblock \showarticletitle{A better world: examples of disability in
  Overwatch}.
\newblock \bibinfo{journal}{\emph{First Person Scholar}}  \bibinfo{volume}{28}
  (\bibinfo{year}{2018}), \bibinfo{pages}{2018}.
\newblock


\bibitem[Davis and Stanovsek(2021)]%
        {davis2021machine}
\bibfield{author}{\bibinfo{person}{Donna~Z Davis} {and} \bibinfo{person}{Shelby
  Stanovsek}.} \bibinfo{year}{2021}\natexlab{}.
\newblock \showarticletitle{The machine as an extension of the body: When
  identity, immersion, and interactive design serve as both resource and
  limitation for the disabled}.
\newblock \bibinfo{journal}{\emph{Human-Machine Communication}}
  \bibinfo{volume}{2} (\bibinfo{year}{2021}), \bibinfo{pages}{121--135}.
\newblock


\bibitem[DiMicco and Millen(2007)]%
        {dimicco2007identity}
\bibfield{author}{\bibinfo{person}{Joan~Morris DiMicco} {and}
  \bibinfo{person}{David~R Millen}.} \bibinfo{year}{2007}\natexlab{}.
\newblock \showarticletitle{Identity management: multiple presentations of self
  in facebook}. In \bibinfo{booktitle}{\emph{Proceedings of the 2007
  international ACM conference on Supporting group work}}.
  \bibinfo{pages}{383--386}.
\newblock


\bibitem[Edwards et~al\mbox{.}(2020)]%
        {edwards2020three}
\bibfield{author}{\bibinfo{person}{Emory~James Edwards},
  \bibinfo{person}{Cella~Monet Sum}, {and} \bibinfo{person}{Stacy~M Branham}.}
  \bibinfo{year}{2020}\natexlab{}.
\newblock \showarticletitle{Three tensions between personas and complex
  disability identities}. In \bibinfo{booktitle}{\emph{Extended abstracts of
  the 2020 CHI conference on human factors in computing systems}}.
  \bibinfo{pages}{1--9}.
\newblock


\bibitem[Freeman and Maloney(2021)]%
        {freeman2021body}
\bibfield{author}{\bibinfo{person}{Guo Freeman} {and} \bibinfo{person}{Divine
  Maloney}.} \bibinfo{year}{2021}\natexlab{}.
\newblock \showarticletitle{Body, avatar, and me: The presentation and
  perception of self in social virtual reality}.
\newblock \bibinfo{journal}{\emph{Proceedings of the ACM on Human-Computer
  Interaction}} \bibinfo{volume}{4}, \bibinfo{number}{CSCW3}
  (\bibinfo{year}{2021}), \bibinfo{pages}{1--27}.
\newblock


\bibitem[Games(2013)]%
        {gtav}
\bibfield{author}{\bibinfo{person}{Rockstar Games}.}
  \bibinfo{year}{2013}\natexlab{}.
\newblock \bibinfo{title}{Grand Theft Auto V}.
\newblock
\newblock


\bibitem[Gerling et~al\mbox{.}(2016)]%
        {gerling2016designing}
\bibfield{author}{\bibinfo{person}{Kathrin Gerling}, \bibinfo{person}{Kieran
  Hicks}, \bibinfo{person}{Michael Kalyn}, \bibinfo{person}{Adam Evans}, {and}
  \bibinfo{person}{Conor Linehan}.} \bibinfo{year}{2016}\natexlab{}.
\newblock \showarticletitle{Designing movement-based play with young people
  using powered wheelchairs}. In \bibinfo{booktitle}{\emph{Proceedings of the
  2016 CHI Conference on Human Factors in Computing Systems}}.
  \bibinfo{pages}{4447--4458}.
\newblock


\bibitem[Goering(2015)]%
        {goering2015rethinking}
\bibfield{author}{\bibinfo{person}{Sara Goering}.}
  \bibinfo{year}{2015}\natexlab{}.
\newblock \showarticletitle{Rethinking disability: the social model of
  disability and chronic disease}.
\newblock \bibinfo{journal}{\emph{Current reviews in musculoskeletal medicine}}
  \bibinfo{volume}{8}, \bibinfo{number}{2} (\bibinfo{year}{2015}),
  \bibinfo{pages}{134--138}.
\newblock


\bibitem[Goffman(1959)]%
        {goffman2021presentation}
\bibfield{author}{\bibinfo{person}{Erving Goffman}.}
  \bibinfo{year}{1959}\natexlab{}.
\newblock \bibinfo{booktitle}{\emph{The presentation of self in everyday
  life}}.
\newblock \bibinfo{publisher}{Bantam Doubleday Dell Publishing group}.
\newblock


\bibitem[Haimson and Hoffmann(2016)]%
        {haimson2016constructing}
\bibfield{author}{\bibinfo{person}{Oliver~L Haimson} {and}
  \bibinfo{person}{Anna~Lauren Hoffmann}.} \bibinfo{year}{2016}\natexlab{}.
\newblock \showarticletitle{Constructing and enforcing" authentic" identity
  online: Facebook, real names, and non-normative identities}.
\newblock \bibinfo{journal}{\emph{First Monday}} (\bibinfo{year}{2016}).
\newblock


\bibitem[Herndon(2002)]%
        {herndon2002disparate}
\bibfield{author}{\bibinfo{person}{April Herndon}.}
  \bibinfo{year}{2002}\natexlab{}.
\newblock \showarticletitle{Disparate but disabled: Fat embodiment and
  disability studies}.
\newblock \bibinfo{journal}{\emph{Nwsa Journal}} (\bibinfo{year}{2002}),
  \bibinfo{pages}{120--137}.
\newblock


\bibitem[Hey(2019)]%
        {hey2019is}
\bibfield{author}{\bibinfo{person}{Alex Hey}.} \bibinfo{year}{2019}\natexlab{}.
\newblock \bibinfo{booktitle}{\emph{Is the "SQUIRREL!" Stereotype True?}}
\newblock
\urldef\tempurl%
\url{https://www.resetadhd.com/blog/squirrel-adhd-stereotype}
\showURL{%
\tempurl}


\bibitem[Higgin(2009)]%
        {higgin2009blackless}
\bibfield{author}{\bibinfo{person}{Tanner Higgin}.}
  \bibinfo{year}{2009}\natexlab{}.
\newblock \showarticletitle{Blackless fantasy: The disappearance of race in
  massively multiplayer online role-playing games}.
\newblock \bibinfo{journal}{\emph{Games and Culture}} \bibinfo{volume}{4},
  \bibinfo{number}{1} (\bibinfo{year}{2009}), \bibinfo{pages}{3--26}.
\newblock


\bibitem[Invalid(2019)]%
        {Sins_Invalid_2019}
\bibfield{author}{\bibinfo{person}{Sins Invalid}.}
  \bibinfo{year}{2019}\natexlab{}.
\newblock \bibinfo{booktitle}{\emph{Skin Tooth and Bone: The Basis of Movement
  is Our People, a Disability Justice Primer} (\bibinfo{edition}{2nd} ed.)}.
\newblock \bibinfo{publisher}{Sins Invalid}.
\newblock


\bibitem[James~Edwards et~al\mbox{.}(2021)]%
        {edwards2021s}
\bibfield{author}{\bibinfo{person}{Emory James~Edwards}, \bibinfo{person}{Kyle
  Lewis~Polster}, \bibinfo{person}{Isabel Tuason}, \bibinfo{person}{Emily
  Blank}, \bibinfo{person}{Michael Gilbert}, {and} \bibinfo{person}{Stacy
  Branham}.} \bibinfo{year}{2021}\natexlab{}.
\newblock \showarticletitle{"That's in the eye of the beholder": Layers of
  Interpretation in Image Descriptions for Fictional Representations of People
  with Disabilities}. In \bibinfo{booktitle}{\emph{The 23rd International ACM
  SIGACCESS Conference on Computers and Accessibility}}.
  \bibinfo{pages}{1--14}.
\newblock


\bibitem[Jin(2009)]%
        {jin2009avatars}
\bibfield{author}{\bibinfo{person}{Seung-A~Annie Jin}.}
  \bibinfo{year}{2009}\natexlab{}.
\newblock \showarticletitle{Avatars mirroring the actual self versus projecting
  the ideal self: The effects of self-priming on interactivity and immersion in
  an exergame, Wii Fit}.
\newblock \bibinfo{journal}{\emph{CyberPsychology \& Behavior}}
  \bibinfo{volume}{12}, \bibinfo{number}{6} (\bibinfo{year}{2009}),
  \bibinfo{pages}{761--765}.
\newblock


\bibitem[Jin and Park(2009)]%
        {jin2009parasocial}
\bibfield{author}{\bibinfo{person}{Seung-A~Annie Jin} {and}
  \bibinfo{person}{Namkee Park}.} \bibinfo{year}{2009}\natexlab{}.
\newblock \showarticletitle{Parasocial interaction with my avatar: Effects of
  interdependent self-construal and the mediating role of self-presence in an
  avatar-based console game, Wii}.
\newblock \bibinfo{journal}{\emph{CyberPsychology \& Behavior}}
  \bibinfo{volume}{12}, \bibinfo{number}{6} (\bibinfo{year}{2009}),
  \bibinfo{pages}{723--727}.
\newblock


\bibitem[Kafai et~al\mbox{.}(2010a)]%
        {kafai2010blacks}
\bibfield{author}{\bibinfo{person}{Yasmin~B Kafai}, \bibinfo{person}{Melissa~S
  Cook}, {and} \bibinfo{person}{Deborah~A Fields}.}
  \bibinfo{year}{2010}\natexlab{a}.
\newblock \showarticletitle{‘‘Blacks Deserve Bodies Too!’’: Design and
  Discussion About Diversity and Race in a Tween Virtual World}.
\newblock \bibinfo{journal}{\emph{Games and Culture}} \bibinfo{volume}{5},
  \bibinfo{number}{1} (\bibinfo{year}{2010}), \bibinfo{pages}{43--63}.
\newblock


\bibitem[Kafai et~al\mbox{.}(2010b)]%
        {kafai2010your}
\bibfield{author}{\bibinfo{person}{Yasmin~B Kafai}, \bibinfo{person}{Deborah~A
  Fields}, {and} \bibinfo{person}{Melissa~S Cook}.}
  \bibinfo{year}{2010}\natexlab{b}.
\newblock \showarticletitle{Your second selves: Player-designed avatars}.
\newblock \bibinfo{journal}{\emph{Games and culture}} \bibinfo{volume}{5},
  \bibinfo{number}{1} (\bibinfo{year}{2010}), \bibinfo{pages}{23--42}.
\newblock


\bibitem[Kai-Cheong~Chan and Gillick(2009)]%
        {kai2009fatness}
\bibfield{author}{\bibinfo{person}{Nathan Kai-Cheong~Chan} {and}
  \bibinfo{person}{Allison~C Gillick}.} \bibinfo{year}{2009}\natexlab{}.
\newblock \showarticletitle{Fatness as a disability: questions of personal and
  group identity}.
\newblock \bibinfo{journal}{\emph{Disability \& Society}} \bibinfo{volume}{24},
  \bibinfo{number}{2} (\bibinfo{year}{2009}), \bibinfo{pages}{231--243}.
\newblock


\bibitem[Kiesler et~al\mbox{.}(1984)]%
        {kiesler1984social}
\bibfield{author}{\bibinfo{person}{Sara Kiesler}, \bibinfo{person}{Jane
  Siegel}, {and} \bibinfo{person}{Timothy~W McGuire}.}
  \bibinfo{year}{1984}\natexlab{}.
\newblock \showarticletitle{Social psychological aspects of computer-mediated
  communication.}
\newblock \bibinfo{journal}{\emph{American psychologist}} \bibinfo{volume}{39},
  \bibinfo{number}{10} (\bibinfo{year}{1984}), \bibinfo{pages}{1123}.
\newblock


\bibitem[Kyr{\"o}l{\"a}(2021)]%
        {kyrola2021fat}
\bibfield{author}{\bibinfo{person}{Katariina Kyr{\"o}l{\"a}}.}
  \bibinfo{year}{2021}\natexlab{}.
\newblock \showarticletitle{Fat in the Media}.
\newblock In \bibinfo{booktitle}{\emph{The Routledge International Handbook of
  Fat Studies}}. \bibinfo{publisher}{Routledge}, \bibinfo{pages}{105--116}.
\newblock


\bibitem[Leonard(2006)]%
        {leonard2006not}
\bibfield{author}{\bibinfo{person}{David~J Leonard}.}
  \bibinfo{year}{2006}\natexlab{}.
\newblock \showarticletitle{Not a hater, just keepin'it real: The importance of
  race-and gender-based game studies}.
\newblock \bibinfo{journal}{\emph{Games and culture}} \bibinfo{volume}{1},
  \bibinfo{number}{1} (\bibinfo{year}{2006}), \bibinfo{pages}{83--88}.
\newblock


\bibitem[Lorde(1982)]%
        {lorde1982learning}
\bibfield{author}{\bibinfo{person}{Audre Lorde}.}
  \bibinfo{year}{1982}\natexlab{}.
\newblock \showarticletitle{Learning from the 60s}.
\newblock \bibinfo{journal}{\emph{Sister outsider: Essays and speeches}}
  \bibinfo{volume}{13444} (\bibinfo{year}{1982}).
\newblock


\bibitem[Mack et~al\mbox{.}(2021)]%
        {mack2021we}
\bibfield{author}{\bibinfo{person}{Kelly Mack}, \bibinfo{person}{Emma
  McDonnell}, \bibinfo{person}{Dhruv Jain}, \bibinfo{person}{Lucy Lu~Wang},
  \bibinfo{person}{Jon E.~Froehlich}, {and} \bibinfo{person}{Leah Findlater}.}
  \bibinfo{year}{2021}\natexlab{}.
\newblock \showarticletitle{What do we mean by “accessibility research”? A
  literature survey of accessibility papers in CHI and ASSETS from 1994 to
  2019}. In \bibinfo{booktitle}{\emph{Proceedings of the 2021 CHI Conference on
  Human Factors in Computing Systems}}. \bibinfo{pages}{1--18}.
\newblock


\bibitem[Mack et~al\mbox{.}(2022)]%
        {mack2022anticipate}
\bibfield{author}{\bibinfo{person}{Kelly Mack}, \bibinfo{person}{Emma
  McDonnell}, \bibinfo{person}{Venkatesh Potluri}, \bibinfo{person}{Maggie Xu},
  \bibinfo{person}{Jailyn Zabala}, \bibinfo{person}{Jeffrey Bigham},
  \bibinfo{person}{Jennifer Mankoff}, {and} \bibinfo{person}{Cynthia Bennett}.}
  \bibinfo{year}{2022}\natexlab{}.
\newblock \showarticletitle{Anticipate and Adjust: Cultivating Access in
  Human-Centered Methods}. In \bibinfo{booktitle}{\emph{CHI Conference on Human
  Factors in Computing Systems}}. \bibinfo{pages}{1--18}.
\newblock


\bibitem[McArthur and Jenson(2014)]%
        {mcarthur2014everyone}
\bibfield{author}{\bibinfo{person}{Victoria McArthur} {and}
  \bibinfo{person}{Jennifer Jenson}.} \bibinfo{year}{2014}\natexlab{}.
\newblock \showarticletitle{E is for everyone? Best practices for the socially
  inclusive design of avatar creation interfaces}. In
  \bibinfo{booktitle}{\emph{Proceedings of the 2014 Conference on Interactive
  Entertainment}}. \bibinfo{pages}{1--8}.
\newblock


\bibitem[McArthur et~al\mbox{.}(2015)]%
        {mcarthur2015avatar}
\bibfield{author}{\bibinfo{person}{Victoria McArthur},
  \bibinfo{person}{Robert~John Teather}, {and} \bibinfo{person}{Jennifer
  Jenson}.} \bibinfo{year}{2015}\natexlab{}.
\newblock \showarticletitle{The avatar affordances framework: mapping
  affordances and design trends in character creation interfaces}. In
  \bibinfo{booktitle}{\emph{Proceedings of the 2015 annual symposium on
  Computer-Human Interaction in Play}}. \bibinfo{pages}{231--240}.
\newblock


\bibitem[Miserandino(2003)]%
        {miserandino2003spoon}
\bibfield{author}{\bibinfo{person}{Christine Miserandino}.}
  \bibinfo{year}{2003}\natexlab{}.
\newblock \bibinfo{booktitle}{\emph{The Spoon Theory written by Christine
  Miserandino}}.
\newblock
\urldef\tempurl%
\url{https://butyoudontlooksick.com/articles/written-by-christine/the-spoon-theory/}
\showURL{%
\tempurl}


\bibitem[Morgan et~al\mbox{.}(2020)]%
        {morgan2020role}
\bibfield{author}{\bibinfo{person}{Helen Morgan}, \bibinfo{person}{Amanda
  O’donovan}, \bibinfo{person}{Renita Almeida}, \bibinfo{person}{Ashleigh
  Lin}, {and} \bibinfo{person}{Yael Perry}.} \bibinfo{year}{2020}\natexlab{}.
\newblock \showarticletitle{The Role of the Avatar in Gaming for Trans and
  Gender Diverse Young People}.
\newblock \bibinfo{journal}{\emph{International journal of environmental
  research and public health}} \bibinfo{volume}{17}, \bibinfo{number}{22}
  (\bibinfo{year}{2020}), \bibinfo{pages}{8617}.
\newblock


\bibitem[Mott et~al\mbox{.}(2019)]%
        {mott2019accessible}
\bibfield{author}{\bibinfo{person}{Martez Mott}, \bibinfo{person}{Edward
  Cutrell}, \bibinfo{person}{Mar~Gonzalez Franco}, \bibinfo{person}{Christian
  Holz}, \bibinfo{person}{Eyal Ofek}, \bibinfo{person}{Richard Stoakley}, {and}
  \bibinfo{person}{Meredith~Ringel Morris}.} \bibinfo{year}{2019}\natexlab{}.
\newblock \showarticletitle{Accessible by design: An opportunity for virtual
  reality}. In \bibinfo{booktitle}{\emph{2019 IEEE International Symposium on
  Mixed and Augmented Reality Adjunct (ISMAR-Adjunct)}}. IEEE,
  \bibinfo{pages}{451--454}.
\newblock


\bibitem[Neustaedter and Fedorovskaya(2009)]%
        {neustaedter2009presenting}
\bibfield{author}{\bibinfo{person}{Carman Neustaedter} {and}
  \bibinfo{person}{Elena~A Fedorovskaya}.} \bibinfo{year}{2009}\natexlab{}.
\newblock \showarticletitle{Presenting identity in a virtual world through
  avatar appearances.}. In \bibinfo{booktitle}{\emph{Graphics Interface}}.
  \bibinfo{pages}{183--190}.
\newblock


\bibitem[Nowak and Fox(2018)]%
        {nowak2018avatars}
\bibfield{author}{\bibinfo{person}{Kristine~L Nowak} {and}
  \bibinfo{person}{Jesse Fox}.} \bibinfo{year}{2018}\natexlab{}.
\newblock \showarticletitle{Avatars and computer-mediated communication: a
  review of the definitions, uses, and effects of digital representations}.
\newblock \bibinfo{journal}{\emph{Review of Communication Research}}
  \bibinfo{volume}{6} (\bibinfo{year}{2018}), \bibinfo{pages}{30--53}.
\newblock


\bibitem[Pace(2008)]%
        {pace2008can}
\bibfield{author}{\bibinfo{person}{Tyler Pace}.}
  \bibinfo{year}{2008}\natexlab{}.
\newblock \showarticletitle{Can an orc catch a cab in stormwind? Cybertype
  preference in the World of Warcraft character creation interface}.
\newblock In \bibinfo{booktitle}{\emph{CHI'08 Extended Abstracts on Human
  Factors in Computing Systems}}. \bibinfo{pages}{2493--2502}.
\newblock


\bibitem[Pace et~al\mbox{.}(2009)]%
        {pace2009socially}
\bibfield{author}{\bibinfo{person}{Tyler Pace}, \bibinfo{person}{Aaron
  Houssian}, {and} \bibinfo{person}{Victoria McArthur}.}
  \bibinfo{year}{2009}\natexlab{}.
\newblock \showarticletitle{Are socially exclusive values embedded in the
  avatar creation interfaces of MMORPGs?}
\newblock \bibinfo{journal}{\emph{Journal of Information, Communication and
  Ethics in Society}} (\bibinfo{year}{2009}).
\newblock


\bibitem[Passmore and Mandryk(2018)]%
        {passmore2018about}
\bibfield{author}{\bibinfo{person}{Cale~J. Passmore} {and}
  \bibinfo{person}{Regan Mandryk}.} \bibinfo{year}{2018}\natexlab{}.
\newblock \showarticletitle{An About Face: Diverse Representation in Games}. In
  \bibinfo{booktitle}{\emph{Proceedings of the 2018 Annual Symposium on
  Computer-Human Interaction in Play}} (Melbourne, VIC, Australia)
  \emph{(\bibinfo{series}{CHI PLAY '18})}. \bibinfo{publisher}{Association for
  Computing Machinery}, \bibinfo{address}{New York, NY, USA},
  \bibinfo{pages}{365–380}.
\newblock
\showISBNx{9781450356244}
\urldef\tempurl%
\url{https://doi.org/10.1145/3242671.3242711}
\showDOI{\tempurl}


\bibitem[Shakespeare(1996)]%
        {shakespeare1996disability}
\bibfield{author}{\bibinfo{person}{Tom Shakespeare}.}
  \bibinfo{year}{1996}\natexlab{}.
\newblock \showarticletitle{Disability, identity and difference}.
\newblock \bibinfo{journal}{\emph{Exploring the divide}}
  (\bibinfo{year}{1996}), \bibinfo{pages}{94--113}.
\newblock


\bibitem[Shakespeare et~al\mbox{.}(2006)]%
        {shakespeare2006social}
\bibfield{author}{\bibinfo{person}{Tom Shakespeare} {et~al\mbox{.}}}
  \bibinfo{year}{2006}\natexlab{}.
\newblock \showarticletitle{The social model of disability}.
\newblock \bibinfo{journal}{\emph{The disability studies reader}}
  \bibinfo{volume}{2} (\bibinfo{year}{2006}), \bibinfo{pages}{197--204}.
\newblock


\bibitem[Turkle(1994)]%
        {turkle1994constructions}
\bibfield{author}{\bibinfo{person}{Sherry Turkle}.}
  \bibinfo{year}{1994}\natexlab{}.
\newblock \showarticletitle{Constructions and reconstructions of self in
  virtual reality: Playing in the MUDs}.
\newblock \bibinfo{journal}{\emph{Mind, Culture, and Activity}}
  \bibinfo{volume}{1}, \bibinfo{number}{3} (\bibinfo{year}{1994}),
  \bibinfo{pages}{158--167}.
\newblock


\bibitem[Turkle(1999)]%
        {turkle1999cyberspace}
\bibfield{author}{\bibinfo{person}{Sherry Turkle}.}
  \bibinfo{year}{1999}\natexlab{}.
\newblock \showarticletitle{Cyberspace and identity}.
\newblock \bibinfo{journal}{\emph{Contemporary sociology}}
  \bibinfo{volume}{28}, \bibinfo{number}{6} (\bibinfo{year}{1999}),
  \bibinfo{pages}{643--648}.
\newblock


\bibitem[Wendell(2001)]%
        {wendell2001unhealthy}
\bibfield{author}{\bibinfo{person}{Susan Wendell}.}
  \bibinfo{year}{2001}\natexlab{}.
\newblock \showarticletitle{Unhealthy disabled: Treating chronic illnesses as
  disabilities}.
\newblock \bibinfo{journal}{\emph{Hypatia}} \bibinfo{volume}{16},
  \bibinfo{number}{4} (\bibinfo{year}{2001}), \bibinfo{pages}{17--33}.
\newblock


\bibitem[Zhang et~al\mbox{.}(2022)]%
        {zhang2022s}
\bibfield{author}{\bibinfo{person}{Kexin Zhang}, \bibinfo{person}{Elmira
  Deldari}, \bibinfo{person}{Zhicong Lu}, \bibinfo{person}{Yaxing Yao}, {and}
  \bibinfo{person}{Yuhang Zhao}.} \bibinfo{year}{2022}\natexlab{}.
\newblock \showarticletitle{``It's Just Part of Me:'' Understanding Avatar
  Diversity and Self-Presentation of People with Disabilities in Social Virtual
  Reality} \emph{(\bibinfo{series}{ASSETS '22})}.
  \bibinfo{publisher}{Association for Computing Machinery},
  \bibinfo{address}{New York, NY, USA}, Article \bibinfo{articleno}{4},
  \bibinfo{numpages}{16}~pages.
\newblock
\showISBNx{9781450392587}
\urldef\tempurl%
\url{https://doi.org/10.1145/3517428.3544829}
\showDOI{\tempurl}


\end{thebibliography}

\appendix

\section{Demographics Summary}

We offer a brief summary of the demographics of our participants.

\textbf{Age}: \textit{M}=29.1 (\textit{SD}=10.5).

\textbf{Gender}: 7 men or trans masc, 6 women, and 5 nonbinary or agender.

\textbf{Race}: 9 White, 5 as Asian or South Asian, 4 as Black or African American, 2 as Latinx or Hispanic, 1 as other, and 3 as multiple/mixed race. 

\textbf{Disability, DHH, chronic or mental illness, or neurodiversity status}: 6 people who are neurodiverse, 6 people with chronic illnesses, 4 DHH people, 4 people with mental illnesses, 3 people with mobility disabilities, 3 BLV people, 2 people who identify as being fat, and 2 people with other disabilities; 10 out of 18 participants had multiple disabilities/conditions.

\end{document}